\documentclass{amsart}

\usepackage{amsmath,amssymb,graphics}
\usepackage[headings]{fullpage}

\usepackage[all]{xy}
\usepackage{graphicx}

\usepackage[all]{xy}

\usepackage{fullpage}

\newtheorem{theorem}{Theorem}[section]

\newtheorem{prop}{Proposition}

\newtheorem{Proposal}{Proposal}

\numberwithin{equation}{section}

\def\l {\lambda}

\renewcommand{\(}{\begin{equation}}
\renewcommand{\)}{\end{equation}}
\newcommand{\bea}{\begin{eqnarray}}
\newcommand{\eea}{\end{eqnarray}}
\newcommand{\R}{{\mathbb R}}
\newcommand{\C}{{\mathbb C}}
\newcommand{\Z}{{\mathbb Z}}

\def\proof {{Proof.}\hspace{7pt}}
\def\endofproof {\hfill{$\Box$}}

\begin{document}

%




\title{Hypermatrix factors for string and membrane junctions}

\author{Yuhan Fang}
\address{Department of Mathematics,
Yale University,
New Haven, CT 06511}
\email{yuhan.fang@yale.edu, hsati@math.umd.edu,
shir.levkowitz@yale.edu, daniel.thompson@yale.edu}

\author{Shir Levkowitz}

\author{Hisham Sati}
\address{Current address (for H. S.): Department of Mathematics,
University of Maryland,
College Park, MD 20742}

\author{Daniel Thompson}

%

\begin{abstract}
The adjoint representations of the Lie algebras 
of the classical groups $SU(n)$, $SO(n)$, and $Sp(n)$
are, respectively, tensor, antisymmetric, and symmetric 
products of two vector spaces, and hence are matrix representations.
We consider the analogous products of three vector spaces 
and study when they appear as summands in Lie algebra 
decompositions. The $\Z_3$-grading of the exceptional 
Lie algebras provides such summands and provides representations 
of classical groups on {\it hypermatrices}. The main 
natural 
application is a formal study of three-junctions of strings and membranes.
Generalizations are also considered. 

\end{abstract}

\maketitle
\tableofcontents


\newpage

\section{Introduction}


Classical Lie groups admit representations on 
vector spaces as well as on 
second powers 
of the vector spaces, underlying the Lie algebra. 
The first is the fundamental representation and the second 
is the adjoint representation. The next step is representations
on triple product of vector spaces, which is the context of this paper.
We consider $\Z_3$-graded decomposition of Lie exceptional 
Lie algebras $\frak{g}$ along the lines of \cite{V1} \cite{V2}. 

\vspace{3mm}
The degree 0 piece is a classical Lie algebra which acts on 
degree 1 and degree -1 pieces via the module structure inherited from 
the Lie bracket on $\frak{g}$. We study this action from the point of view
of representations on {\it hypermatrices}, which are higher-dimensional
generalizations of matrices. The latter are two-dimensional arrays of numbers
while the former are $n$-dimensional such arrays, $n \geq 0$.   
These can take values
  in $\C$ or $\R$ or even finite fields. On these, we will use scalar invariants,
  the corresponding trace and hyperdeterminant \cite{GKZ} generalizing the
  usual trace and determinant of matrices. 
Generally, the situations we encounter are summarized as follows
 $$
\{{\rm Exceptional~ Lie ~algebra}\}= \{{\rm ``dual" ~cubic ~hypermatrix~}\}
\oplus \{{\rm classical ~Lie ~algebra~}\} \oplus \{{\rm ~cubic~ hypermatrix}\}\;. 
$$
All exceptional algebras appear, most notably $E_8$ and $E_6$, corresponding 
to summands the tensor power $\otimes^3 V$ and the exterior 
power $\wedge^3 V$ in the graded 
Lie algebra decomposition, as well as their subgroups such as $D_4$,
which corresponds to a summand the symmetric power $S^3 V$.

\vspace{3mm}
Hyperdeterminants have appeared in applications to string theory,
starting in \cite{D} (see \cite{BDDER} for a review). We consider 
other applications, where not only hyperdeterminants but also 
hypermatrices also appear, in the following context. 
One of the original motivations for  string theory 
was to describe mesons. 
 A meson is formed of a quark $q$ and an antiquark
 $\overline{q}$, i.e.
$q\overline{q}$.
 The modern viewpoint (see \cite{Pol} for details) is that the 
endpoints of the strings carry $U(1)$ degrees of freedom
and can end on D-branes.
The gauge group arising from $n$ coincident D-branes
becomes nonabelian $U(n)$. The $U(1)$ corresponds to
degrees of freedom for 
 the center of mass and the $SU(n)$ for the relative degrees
 of freedom. 
There are also models that extend 
the above description to baryons (see e.g. \cite{KN}). 
A baryon is formed of a triplet of quarks, 
i.e. $qqq$. 
 The modern incarnation of this is string junctions
or prongs \cite{Sch}.

\vspace{3mm}
 Since the prongs of a three-pronged string 
 are mutually non-local, they 
cannot all end on D-branes in general. 
The exception  is 
the D3-brane, on which any $(p,q)$ string can end \cite{B}.
Here $(p,q)$ denotes an $SL(2,\Z)$ doublet 
with $p$ and $q$ coprimes integers. 
Note that by S-duality one can have $(p,q)$ strings
and $(p,q)$ D3-branes. 
Thus D3-branes are allowed boundaries for three-pronged 
open strings. Since one needs at least three D3-branes to 
support a three-pronged string, the states should arise for 
gauge groups at least as large as $SU(3)$ \cite{B}.
D7-branes allow for gauge groups other than 
$SU(n)$, namely $S{\rm O}(n)$, $E_6$, $E_7$, and $E_8$
\cite{Jo} \cite{GZ}.


 \vspace{3mm}
 In gauge theory, the number of degrees of freedom corresponding 
to $U(n)$ is $n^2$, which is the dimension of the adjoint representation.
This appears for theory of the open string ending on $n$ coincident D-branes. 
On the other hand, the membrane in M-theory can end on the fivebrane
\cite{St}.
This M-brane configuration is T-dual to the above-mentioned
picture of having 
strings end on D3-branes. 
One can consider open membranes ending on multiple fivebranes in analogy 
to open strings ending on multiple D-branes. 
The triple string junction
arises from M-theory by starting with a pant configuration of membrane and wrapping each
of the membrane prongs on different cycles of the compactified two-dimensional torus
\cite{ASY} \cite{Sch}.
The resulting field theory is not well 
understood. The number of degrees of 
freedom in this case scale as the cube $n^3$ of the number of M5-branes
 \cite{KT} \cite{HS} \cite{BFT}. 
This suggests that a description might fall outside the scope of
  finite $n$-dimensional semi-simple Lie groups and algebras
  as none of those have a dimension growing as fast as $n^3$ 
  (where $n$ would be the dimension
  of the Cartan sub-algebra) \cite{KT} \cite{BHS1} \cite{BHS2}.

\vspace{3mm}
Configurations with multiple membranes are also allowed. 
The membrane fields do not have to be in the Lie algebra of $N \times N$ matrices
${\rm Mat}_N(\C)$ and
the membrane five-brane interaction seems to be out of the realm of
matrix theory at the moment \cite{B}.
It is shown in \cite{DFMR} that the Lie 3-algebras proposed in 
\cite{BL} to model multiple membranes can be encoded in 
an `ordinary' Lie algebra together with some representation. 
In fact, the relation found in reference \cite{DFMR} between
classes of metric 3-algebras and unitary representations of Lie algebras is
much more general than for just the 3-Lie algebras, which only encode
maximally supersymmetric M2-brane theories. For example, it applies 
to the ABJM theory in
reference \cite{ABJM}, where the 3-algebra corresponds to a so-called anti-Jordan
triple
system rather than a Lie 3-algebra. The precise relationship was clarified
in \cite{DFM} for all the M2-brane theories which are at least
half-BPS.
Hence, Lie 3-algebras do not seem to be absolutely indispensable for
models of multiple membranes, e.g. \cite{ABJM}.
Therefore,
in this note we propose to keep working with Lie algebras, but to 
view them from a different angle as above. 

\vspace{3mm}
The representations we consider are
not the fundamental.
Other representations, which are direct sums,
were considered in 
\cite{G2} to implement an exceptional symmetry, namely that
of the real Lie algebra of type $G_2$. The complex case, $G_2(\C)$ 
cannot be seen within the Lie 3-algebra formalism since in this case
the vector space $V=\C^3$ has dimension 3, and hence cannot support 
(complex) Lie 3-algebras. However, the complex case can be 
implemented in the current context. 
The implementation of the above
proposal leads further to exceptional algebras of type $E$ and $F$, as
well as $\frak{so}(8)$. 

\vspace{3mm}
As a natural byproduct of our formalism, we show that the symmetry of the 
gauge fields resulting from the dimensional reduction of eleven-dimensional 
supergravity to three dimensions is that of the exceptional Lie algebra
$\frak{e}_8$. This is obtained in section \ref{sugra}
  using a $\Z_3$-graded model for $\frak{e}_8$, and thus proves 
  an assertion in \cite{CJLP}.

\vspace{3mm}
To make the paper as self-contained
as possible, we have kept enough expository parts both on elementary-- but perhaps not 
widely known -- discussions of 
 nonlinear algebra and representation theory, as well as 
on the applications to strings and branes in physics.
  
\section{Tensor Product Decompositions and Lie Algebras} 

\noindent{\bf Reminder on Lie algebra representations.} We start by reviewing 
some basic notions which we will use in this paper. 
\begin{enumerate}
\item A vector space $W$ is called a {\it representation} of a Lie algebra $\frak{g}$,
or a $\frak{g}$-module, if there is a Lie algebra homomorphism $\frak{g} \to \frak{gl}(W)$.
\item When $W=\frak{g}$, the map ${\rm ad}_{\frak{g}}: \frak{g} \to \frak{gl}(\frak{g})$ defined by 
${\rm ad}_X (Y)=[X,Y]$, the Lie bracket of $X, Y \in \frak{g}$, is the {\it adjoint representation}
of $\frak{g}$. 
\item Let $V$ be a $\frak{g}$-module. Then $W^*$ is the {\it dual} (or {\it contragredient})
representation given by $(X \cdot f) (v)=-f (X \cdot v)$ for $X \in \frak{g}$, $f \in W^*$, $v \in W$. 
\item The dual of the adjoint representation is the {\it coadjoint} representation of $\frak{g}$ 
on $\frak{g}^*$ is a map ${\rm ad}^*: \frak{g} \to \frak{gl}(\frak{g}^*)$ defined by 
${\rm ad}_X^* a (Y) = a (- {\rm ad}_XY)=-a\left([X, Y] \right)$, for $a \in \frak{g}^*$,
$X, Y \in \frak{g}$.

\end{enumerate}

\subsection{The case of an open string}


The open string Chan-Paton \cite{CP} factors lead to matrix Lie
groups as follows (see \cite{GSW}):
\begin{enumerate}
\item Assign a vector space $V$ to each of the two
point boundaries of the open string.
\item Form the tensor product $V \otimes V$ in the case of unoriented string and
$V \otimes \overline{V}$, where $\overline{V}$ is the complex conjugate, in the
case of oriented strings. The former tensor product is a special case of the latter
when $\overline{V}=V$. 
\item Explicitly, the states for the two-ended open string 
are represented by matrices
$\lambda^i{}_j$, where $i$ is an index for the states of a `quark' and 
$j$ is an index for the states of the corresponding ``antiquark". 

\item Require that the adjoint representation ${\rm ad} \frak{g}$ 
be (inside) $V \otimes \overline{V}$ so that the spectrum of the string contains a vector gauge field. 
\item The set of anti-hermitian operators is required to form an algebra, as well
as the set of linear combinations of hermitian and anti-hermitian operators.
This means that $\frak{g}=\frak{g}_a \oplus \frak{g}_h$, with
$\frak{g}_a$ required to be a Lie algebra. By a theorem of Wedderburn,
the algebra $\frak{g}$ corresponds to the group $GL(n, \C)$, whose anti-hermitian 
part corresponds to the group $U(n)$. Taking a real form first then the anti-hermitian part gives
two cases: the orthogonal group $SO(n)$ and the symplectic group $Sp(n)$.
Thus the following cases are realized 
\bea
U(n)&:& \qquad {\rm ad} \frak{g}\cong V \otimes \overline{V}\;,
\nonumber\\
SO(n)&:& \qquad {\rm ad} \frak{g}\cong \wedge^2V\;, 
\nonumber\\
Sp(n)&:& \qquad {\rm ad} \frak{g}\cong S^2 V\;.
\label{gsw list}
\eea

\end{enumerate}

\vspace{3mm}
For any finite-dimensional vector space $V$ there is a decomposition 
of $V \otimes V$, under the action of $GL(V)$,
 into a direct sum of irreducible $GL(V)$-modules
\(
V \otimes V = \wedge^2 V \oplus S^2 V\;.
\label{2 decomp}
\)
This means that the above three cases correspond, respectively, to 
the left hand side, to the first summand, and to the second summand in
(\ref{2 decomp}). 

\vspace{3mm}
\noindent {\bf Remark.}
Note that in the complex case the amplitude is invariant under $GL(n, \C)$,
while insisting on the norm of the states to be invariant requires $U(n)$
(see \cite{Pol}). Similarly for the real and quaternionic cases.

\subsection{The case of a junction}
\label{case junction}
We would like to carry out the corresponding process for 
the three-junction. We proceed as follows
\begin{enumerate}
\item We assign a vector space $V_i$, $i=1,2,3$, to each of the three
vertices.

\item We form the tensor product $V_1 \otimes V_2 \otimes V_3$. Then we identify this
with a representation of some Lie (or Kac-Moody) group. If this is not possible then 
identify a summand of this triple tensor product  with a representation of a group.
If we require to have a field in string theory or in M-theory 
to be included in the spectrum, then for the latter an obvious choice would be 
a three-form corresponding to the $C$-field. But we will not insist on this.


\item The action of $GL(V)$ breaks $V \otimes V \otimes V$ into a direct sum 
of four $GL(V)$-modules 
\(
V \otimes V \otimes V = \wedge^3 V \oplus S^3 V \oplus 
\left( \mathbb{S}_{(2,1)} V\right)^{\oplus 2}\;,
\label{triple}
\)
where $\mathbb{S}_{(2,1)} V$ is defined as (see e.g. \cite{FH})
\(
\mathbb{S}_{(2,1)} V=\ker \left( \wedge^2 V \otimes V \longrightarrow 
\wedge^3 V\right)\;.
\label{s21}
\)
Elements of $ \wedge^2 V \otimes V$ are of the form $(v_1 \wedge v_3)\otimes v_2$,
and are embedded in $\wedge^3 V$ as 
$v_1 \otimes v_2 \otimes v_3 - v_3 \otimes v_2 \otimes v_1$. 


\item We thus ask for the summands in the triple tensor product 
(\ref{triple}) to be representation spaces for Lie groups or Lie algebras--
as they cannot be Lie algebras by themselves-- so as to give a `higher analog'
 of the
adjoint representation. 

\item We could also ask for a summand in the graded decomposition of the 
Lie algebra $\frak{g}$ (cf. section \ref{gla}) to be 
identified with a summand in $V \otimes V \otimes V$. 

\item 
The states for the 3-junction are represented by higher matrices $\lambda_{ijk}$,
where each of the indices represents a state of a quark. 
The study of this $\lambda_{ijk}$ is the main subject of this paper.

\end{enumerate}

\vspace{3mm}
A representation `with three indices' mentioned in the introduction 
  should correspond to a product of three
  vector representations, each corresponding to a vector space 
  $V_i$, $i=1,2,3$.  
  There are three possibilities: 
  \begin{enumerate}
  \item Tensor product: $V_1 \otimes V_2 \otimes V_3$
  \item Symmetric power: $S^3 V$, where $V$ is isomorphic to each of the 
  $V_i$. 
 \item Antisymmetric power: $\wedge^3 V$, where again $V$ is 
 isomorphic to each of the  $V_i$. 

\end{enumerate}
  
\vspace{3mm} 
In the desired cases, the grading naturally provides an action of the general
  (or special) linear group on $\wedge^3V$, $\otimes^3V$, or $S^3V$.
  The dimensions all grow $\sim n^3$
\bea
  \dim S^3 (V) &=& 
\frac{1}{6}n (n+1)(n+2)
\;,
\\
  \dim \otimes^3 (V) &=& 
n^3
\;,
\\
  \dim \wedge^3 (V) &=& 
\frac{1}{6}n(n-1) (n-2)
\;.
\eea

\vspace{3mm}
 The question now is what replaces the list (\ref{gsw list}) in the degree 
  three case? We will answer this in section \ref{gla}. It turns out that they 
  correspond not to classical Lie groups but to {\it exceptional} Lie groups!
  
  \vspace{3mm}
  The factors  $\lambda_{ijk}$ a priori admit	 no symmetry, i.e. belong to
   $ V \otimes V \otimes V$.
 If we require antisymmetry upon exchange of the first to indices
 $\lambda_{ijk}=-\lambda_{jik}$ then $\lambda \in \wedge^2 V \otimes V$.
 Using the decomposition 
 \(
 \wedge^2 V \otimes V =\wedge^3 V \oplus
 {\mathbb{S}}_{(2,1)}(V)\;.
 \)
this gives two types for $\lambda$:
\begin{enumerate}
\item $\lambda_{ijk}\in \wedge^3 V$ totally skew-symmetric,
\item $\lambda_{ijk} \in \mathbb{S}_{(2,1)}(V)$ which is such that
\(
\lambda_{ijk} + \lambda_{kij} + \lambda_{jki}=0\;.
\label{trip}
\)


\noindent Furthermore, 
\item if 
$
\lambda_{ijk}=\lambda_{jik}=\lambda_{ikj}=\lambda_{kij}=\lambda_{jki}=\lambda_{kji}
$,
 then $\lambda\in S^3 V$. 
 \end{enumerate}

\vspace{3mm}
\noindent {\bf Remarks.}
\noindent {\bf 1.} The above procedure can be performed on the dual vector space
$V^*$ leading to factors $\wedge^3 V^*$, $\otimes^3 V^*$, and 
$S^3 V^*$, with corresponding factors $\lambda^{*ijk}$. The forms related
to the dual vector space are contravariant while those related to the initial 
vector space are covariant. The duality between $V$ (and its powers)
and $V^*$ (and its powers) is occurring as a duality {\it on} the brane. 
\\
\noindent {\bf 2.}  An alternative considered in \cite{BC} 
 is the fuzzy 3-sphere algebra $\mathcal{A}_n(S^3)$, which
  reduces to the classical algebra of functions
on the 3-sphere in the large $N$ limit. This algebra is not closed under multiplication and so 
a projection is needed after multiplication.
\footnote{The closedness in our context is considered at the end of section
\ref{gla}.}
 This leads to a nonassociative algebra.
 The number of degrees of freedom is given by $D = \frac{1}{6}(n + 1)(n + 2)(2n + 3)$ so that 
 in the large $n\sim \sqrt{N}$ limit this scales as
 $D \sim N^{3/2}$.


\subsection{Graded Lie algebras}
\label{gla}
\begin{enumerate}
\item A Lie algebra $\frak{g}$ is called the {\it direct sum} of two Lie subalgebras 
$\frak{g}=\frak{g}_1 \oplus \frak{g}_2$ if the underlying vector spaces 
obey the direct sum with 
\(
\frak{g}_1\cap \frak{g}_2 =\emptyset;, \qquad \qquad [\frak{g}_1, \frak{g}_2]=0\;.
\label{ideal}
\)
So both $\frak{g}_1$ and $\frak{g}_2$ are ideals of the direct sum. 
\item A Lie algebra $\frak{g}$ is called a {\it semidirect sum} of two Lie subalgebras
$\frak{g}=\frak{g}_1 \oplus_s \frak{g}_2$ if we replace the second condition in 
(\ref{ideal}) by $[\frak{g}_1, \frak{g}_2] \subset \frak{g}_1$, so that $\frak{g}_1$ is an 
ideal but $\frak{g}_2$ is not.
\end{enumerate}

\vspace{3mm}
If $\frak{g}$ is a Lie algebra then the tensor product 
space $\C \otimes \frak{g}$ is a complex vector space since we can define
\(
\tau (\mu \otimes x)= (\tau \mu) \otimes x\;,~~~
\forall~ \tau, \mu \in \C~~~{\rm and~all}~ x \in \frak{g}. 
\)
This can be regarded as a complex Lie algebra $\frak{g}_\C$, 
the complexification 
of $\frak{g}$, if we set for the Lie bracket
\(
\left[ \tau \otimes x~,~ \mu \otimes y \right]=
(\tau \mu) \otimes [x~, ~y]\;,
\)
as then this would still satisfy antisymmetry and 
the Jacobi identity. 

\vspace{3mm}
A graded Lie algebra is an ordinary Lie algebra $\frak{g}$, together with 
a gradation of vector spaces
\(
\frak{g}=\bigoplus_{i \in \Z~{\rm or~} \Z_m} \frak{g}_i\;,
\)
such that the Lie bracket respects this gradation 
\(
\left[\frak{g}_i~,~ \frak{g}_j \right] \subseteq \frak{g}_{i+j}\;.
\)
\begin{itemize}
\item A $\Z_2$-grading $\frak{g}=\frak{g}_0 \oplus \frak{g}_1$ corresponds to 
coset spaces. 

\item A $\Z_3$-grading is of the form
$
\frak{g}=\frak{g}_{-1} \oplus \frak{g}_0 \oplus \frak{g}_1
$.
\item A $\Z$-grading is of the form
$
\frak{g}=\frak{g}_{-d}\oplus \cdots \oplus \frak{g}_{-1} \oplus 
\frak{g}_0 \oplus \frak{g}_1 \oplus \cdots \oplus
\frak{g}_d
$,
where $d=\max \{ p ~|~\frak{g}_p \neq 0\}$ is the {\it depth} of the grading. 

\end{itemize}

\noindent{\bf The grading via the Weyl group.}
Vinberg \cite{V1} extended the concept of Weyl group $W$ to semisimple complex
Lie algebras which are graded modulo any $m$. $W$ is generated by complex reflections,
i.e. linear transformations that can be described in some basis by a matrix of the 
form 
\(
\left(
\begin{array}{cccc}
\omega &&&\\
&1&&\\
&&\ddots&\\
&&&1
\end{array}
\right)\;,
\)
where $\omega$ is a root of unity. 
If $\frak{g}$ is  $\Z_m$-graded for finite $m$ then the linear transformation
$d\theta$ defined by
$d\theta (x)= \omega^k x$, for $x \in \frak{g}_k$, gives the gradation so that $\frak{g}_k$
are the eigenspaces of $d\theta$ as follows \cite{V1}.
For any $\tau \in \C$, set 
\(
\frak{g}(\tau)= \left\{ x \in \frak{g} ~|~ d\theta (x)=\tau x \right\}\;,
\)
so that $\frak{g}=\bigoplus_{\tau} \frak{g}(\tau)$, and 
\(
[\frak{g}(\tau)~,~ \frak{g}(\omega)] \subset \frak{g}(\tau \omega)\;.
\)
The eigenvalues of the operator $d\theta$ can be assumed, without loss
of generality, to be of the form $\omega^k$, with $k\in \Z$. 
Setting $\frak{g}(\omega^k)=\frak{g}_k$ gives a $\Z$-grading of 
$\frak{g}$ if $\theta$ has infinite order, and a $\Z_m$-grading if $\theta$ has
finite order $m$.


\vspace{3mm}
\noindent {\bf Tensor representations of Lie algebras.}
Let $W$ be a $\frak{g}$-module and let $T$, $S$ and $\wedge$ denote 
tensor, symmetric, and antisymmetric powers, respectively.  Then
\begin{enumerate}
\item $T(W)=\bigoplus_{i=0}^{\infty} T^i (W)$ (or $T^i(W)$) is the {\it tensor product representation} of 
$\frak{g}$. 
\item $S(W)=\bigoplus_{i=0}^{\infty} S^i (W)$ (or $S^i(W)$) is the {\it symmetric product representation} of 
$\frak{g}$. 
\item$\wedge (W)= \bigoplus_{i=0}^{\infty} \wedge^i (W)$ (or $\wedge^i (W)$) is the {\it antisymmetric
 product representation} of $\frak{g}$. 
\end{enumerate}

\vspace{3mm}
 As also mentioned in the introduction, there are no Lie groups or algebras whose dimension grows like
  the cube of their rank.  Therefore one cannot find a representation
  of dimension $n^3$ to make up a whole of a Lie algebra. However, the
  next best thing one could hope for is to find inside a Lie algebra a
  representation that grows like $n^3$.
 Thus we seek those Lie algebras $\mathfrak{g}$ which admit a
  decomposition of the form \bea \mathfrak{g}&\supset& {\otimes}^3
  V\;, ~{\rm or}
  \\
  \mathfrak{g}&\supset& {\wedge}^3 V\;, ~{\rm or}
  \\
  \mathfrak{g}&\supset& S^3 V\;.  \eea
Similar requirements can be made for the dual vector spaces
$\otimes^3 V^*$, $\wedge^3 V^*$, and 
$S^3V^*$.
  It turns out that the above decompositions are realized for the Lie algebras
  $\mathfrak{e} _{6}:={\rm Lie}(E_6)$, $\mathfrak{e} _{8}:={\rm Lie}(E_8)$, 
  and $\mathfrak{d} _{4}:={\rm Lie}(D_4)$,
  respectively. From \cite{V2} we have
%

\begin{prop}
Consider the decomposition $\frak{g}=\frak{g}_{-1} \oplus \frak{g}_0 
\oplus \frak{g}_{1}$, where $\frak{g}_0$ is of type  $\frak{sl}$ or $\frak{gl}$,
and $\frak{g}_{-1}$ and $\frak{g}_1$ are third tensor, symmetric, or
antisymmetric powers of some vector space $V$ or the tensor product of 
three vector spaces $V_1, V_2, V_2$. 
The only possibilities 
are
\begin{enumerate}
\item $\frak{d}_4= S^3 V^* \oplus \frak{sl}(V) \oplus S^3 V$, \hspace{3mm} $\dim (V)=3$.
\item $\frak{e}_6=(V_1^* \otimes V_2^* \otimes V_3^*) 
\oplus
\left( \frak{sl}(V_1) \oplus \frak{sl}(V_2) \oplus \frak{sl}(V_3) \right)
\oplus (V_1 \otimes V_2 \otimes V_3)$, \hspace{3mm}  $\dim (V)=3$.
\item $\frak{e}_8=\wedge^3 V^* \oplus \frak{sl}(V) \oplus \wedge^3 V$,
\hspace{3mm}  $\dim (V)=9$. 
\end{enumerate}
\label{prop case}
\end{prop}

%
  
  \vspace{3mm}
  \noindent {\bf Remarks.}
  {\bf 1.} In proposition \ref{prop case}, we think of $V$ as $\C^3$ 
  in (1) and as $\C^9$ in (3), while we think of of $V_i$, $i=1,2,3$, as 
  $\C^3$ in (2).
  
\noindent  {\bf 2.} While in the open string case $\frak{g}_a$ was a Lie algebra, 
  in the three-junction case $\otimes^3$, $\wedge^3V$ and 
  $S^3V$ are not algebras, but only modules. However, in one 
  model they close in the $\frak{g}$-summand $\wedge^3V^*$ (see equation  \ref{cc}) and
  in another they close in the $\frak{g}$-summand 
  $\wedge^6 V$ (cf. equation   \ref{lc}).

\subsection{Representations of the corresponding groups}
\label{rep groups}

 Let $G$ be a connected reductive algebraic group over $\C$ and let $\frak{g}$
be its Lie algebra. Let $\theta$ be a semisimple automorphism of $G$. 
This is the `antiderivative' of $d\theta$, the automorphism of the algebras 
considered in the previous section.

\vspace{3mm}
 Let $G_0$ be the identity component of the group 
$G^{\theta}$ of elements invariant under $\theta$. The two coincide if $G$ 
is simply connected and semisimple. Let $\hat{G}_0$ be the simply 
connected group locally isomorphic to $G_0$. 
The adjoint representation of $G$ induces a linear representation of $G_0$
in each of the subspaces $\frak{g}(\tau)$. 
  The algebra of invariant polynomials $\C [\frak{g}_1]^{G_0}$ is finitely 
  generated and free \cite{V2}.
  
  \vspace{3mm}
  We seek $G_0$-invariant rank 3 tensors.
  For $m=3$ there are the following cases corresponding to the ones in 
  proposition \ref{prop case}
  
  \begin{prop}
  Three-junctions (with no physical constraints) may admit the following 
  group symmetries 
 
 \noindent 1. $G=D_4$, $G_0=SL(3)$, and the elements of $\frak{g}_1$ are 
 symmetric forms of 
  degree three in three variables. 
  
  \noindent 2. $G=E_6$, $\hat{G}_0=SL(3) \times SL(3)\times SL(3)$, and 
  $\frak{g}_1$ can be interpreted as $\C^3 \times \C^3 \times \C^3$. 
  
  \noindent 3. $G=E_8$, $G_0=SL(9)$, and the space $\frak{g}_1$ can be 
  interpreted as the third exterior power $\wedge^3 \C^9$ of $\C^9$.  
  \end{prop}
 
  \vspace{3mm}
  Lie groups and Lie algebras have natural representations, e.g. the
  adjoint, on matrices. Given the above decompositions containing cube powers, 
  corresponding to tensor representations, 
  it is natural to ask what are the corresponding objects replacing matrices.
  The answer is {\it hypermatrices}. What replaces linear algebra is 
  {\it multi-linear algebra}.

  \section{Tensors and Hypermatrices}

\subsection{Hypermatrices and hyperdeterminants}
\label{hh}
A 3-dimensional hypermatrix is a 3-way array of complex numbers 
$A=[a_{j_1j_2 j_3}]_{j_1, j_2, j_3=1}^{n_1, n_2, n_3}$,
where $a_{j_1 j_2 j_3} \in \C$ is the $(j_1, j_2, j_3)$-entry
of the array, and the notation $[\cdot]_{j_1, j_2, j_3=1}^{n_1, n_2, n_3}$ means that the
indices $j_i$ run as  $1\leq j_i \leq n_i$, for $i=1,2,3$. 
This array is denoted as $\C^{n_1 \times n_2 \times n_3}$, 
which is a complex vector space of dimension $n_1n_2n_3$. 

\vspace{3mm}
In general, hypermatrices are higher-dimensional arrays generalizing 
matrices, which are viewed as two-dimensional arrays of numbers. 
The latter admit scalar invariants which include the determinant,
and likewise the former admits the hyperdeterminant. 
For a $k$-dimensional hypermatrix $A=(A_{i_1, \cdots, i_k})_{1\leq i_1, \cdots, i_k \leq n}$ 
of order $n$, the 
{\it hyperdeterminant} of $A$ is
\(
{\mathcal D}{\rm et}_k (A)=\frac{1}{n!} \sum_{\sigma_1, \cdots, \sigma_k \in \Sigma_n}
{\rm sign}(\sigma_1, \cdots, \sigma_k) \prod_{i=1}^n A_{\sigma_1(i)\cdots \sigma_k(i)}\;.
\label{hyperdet}
\)
\begin{enumerate}
\item When $k=2$ this expression for the hyperdeterminant coincide with that of the determinant 
\(
{\mathcal D}{\rm et}_2 (A)={\rm det} (A)= \frac{1}{n!} \sum_{\sigma_1, \sigma_2\in \Sigma_n}
{\rm sign}(\sigma_1){\rm sign}(\sigma_2) \prod_{i=1}^n A_{\sigma_1(i)\sigma_2(i)}\;.
\)
\item Expression (\ref{hyperdet}) is the zero polynomial when the dimension $k$ of the hypermatrix is 
odd. This will be used later in section \ref{GBI} , where an extension to 
the odd-dimensional case is considered.  
\end{enumerate}
\vspace{3mm}
In the $n$-dimensional case, rows and columns are replaced by slices
which come in $n$ types. For example, for $n=3$ we have vertical,
horizontal and lateral slices.
Row and column
operations are replaced by {\it slab} (or {\it slice}) {\it operations} and hence it is natural 
to check for behavior of hypermatrices under those, that is to check analogs for
hypermatrices of Gaussian
elimination for matrices. The following
are essentially known since Cayley \cite{Cay} (see \cite{So} for a more
recent reference). 

\vspace{3mm}
\noindent {\bf Properties of hyperdeterminant under hypermatrix operations.}
\begin{itemize}
  \item[(a)] Interchanging two parallel slices leaves the
    hyperdeterminant invariant up to sign (which may equal 1).
  \item[(b)] The hyperdeterminant is a homogeneous polynomial in the
    entries of each slice.  The degree of homogeneity is the same for
    parallel slices.
  \item[(c)] The hyperdeterminant does not change if we add to some
    slice a scalar multiple of a parallel slice.
  \item[(d)] The hyperdeterminant of a matrix having two parallel
    slices proportional to each other is equal to 0.  In particular,
    $\mathrm{Det}(A)=0$ if $A$ has a zero slice.
\end{itemize} 

\subsection{Equivalence of tensors and hypermatrices}
A 3-array can be formed out of 3 vectors as follows. 
The Segre product of 3 vectors $u \in \C^{n_1}$,
$v \in \C^{n_2}$, and $w \in \C^{n_3}$, is defined as
\(
u \otimes v \otimes w :=[u_{j_1} v_{j_2} w_{j_3}]_{j_1, j_2, j_3=1}^{n_1, n_2, n_3}\;.
\)
Next, for arrays themselves we have that 
the outer product of two 3-arrays $A$ and $B$ is a 6-array
$C=A \otimes B$ with entries
\(
c_{i_1 i_2 i_3j_1 j_2j_3}:= a_{i_1 i_2 i_3} b_{j_1 j_2 j_3}\;.
\)

\vspace{3mm}
\noindent {\bf  The relation of a hypermatrix to a tensor.}
A tensor is an element of 
in the tensor product of vector spaces. The Segre map 
\bea
\varphi : \C^{n_1} \times \C^{n_2} \times \C^{n_k} 
&\longrightarrow & \C^{n_1 \times n_2 \times n_3}
\nonumber\\
(u~,~ v~,~ w) ~~~~~&\longmapsto & u \otimes v \otimes w
\eea
is multilinear with kernel the {\it decomposable tensors}, i.e. those that are 
of the form $A=e_{i_1} \otimes e_{i_2} \otimes e_{i_3}$. 
By the universal property of the tensor product there exists 
a linear map $\theta$
\(
\xymatrix{
 && 
 \C^{n_1} \otimes \C^{n_2} \otimes \C^{n_3}
 \ar[d]^{\theta}
 \\
 \C^{n_1} \times \C^{n_2} \times \C^{n_3}
 \ar[urr]
 \ar[rr]^{\varphi}
 &&
 \C^{n_1 \times n_2 \times n_3}
}
\;.
\)
Since the spaces have the same dimension, $\theta$ is an
isomorphism of vector spaces. Consider the canonical basis
of 
$\C^{n_1} \otimes \C^{n_2} \otimes \C^{n_3}$
\(
\left\{ e_{j_1}^{(1)}\otimes e_{j_2}^{(2)} \otimes e_{j_3}^{(3)} ~|~
1 \leq j_1 \leq n_1, ~1\leq j_2 \leq n_2,~ 1\leq j_3 \leq n_3 \right\}\;,
\)
where $\{e_{n_1}^{(\ell)}, \cdots, e_{n_{\ell}}^{(\ell)} \}$
denotes the canonical basis in $\C^{n_{\ell}}$, $\ell = 1, 2, 3$. Then $\theta$
may be described as \cite{CGLM}
\(
\theta \left(
\sum_{j_1, j_2, j_3}^{n_1, n_2, n_3} a_{j_1, j_2, j_3} e_{j_1}^{(1)} \otimes
e_{j_2}^{(2)} \otimes e_{j_3}^{(3)}
\right)
=
[a_{j_1 j_2 j_3}]_{j_1, j_2, j_3=1}^{n_1, n_2, n_3}\;.
\)
Thus, we have
\begin{prop}
An order 3-tensor in $\C^{n_1} \otimes \C^{n_2} \otimes \C^{n_3}$
is the same as a 3-dimensional hypermatrix in $\C^{n_1 \times n_2 \times n_3}$ 
in the above basis. Similarly for the real case. 
\end{prop}

\subsection{Relation to matrices}

\noindent  {\bf Change of basis.}
Let $A=[a_{ijk}] \in \C^{n_1 \times n_2 \times n_3}$ and let 
$L$, $M$, and $N$ be three $n_1 \times n_1$, $n_2 \times n_2$, and 
$n_3 \times n_3$ nonsingular matrices, respectively. This means that
$L=[l_{ij}] \in GL(n_1, \C)$, $M=[m_{ij}] \in GL(n_2, \C)$, and 
$N=[n_{ij}]\in GL(n_3,\C)$.
The result of the transformation 
of the multilinear map $(L,M,N)$ on $A$ is a tensor $A'=(L,M,N)\cdot A=[a'_{pqr}]\in
\C^{n_1 \times n_2 \times n_3}$, the multilinear transform of $A$, 
defined by 
\(
a'_{pqr}= \sum_{i,j,k} l_{pi} \hspace{0.5mm} m_{qj} \hspace{0.5mm}  n_{rk}
\hspace{0.5mm}  a_{ijk}\;.
\)

\vspace{3mm}
\noindent {\bf Multilinear matrix multiplication.}
The following properties hold (see \cite{DL})
\begin{enumerate}
\item For $A, B \in \C^{n_1 \times n_2 \times n_3}$, $L_i \in GL(n_i, \C)$, 
 and $\alpha, \beta \in \C$,
 $$
 (L_1, L_2, L_3)\cdot (\alpha A + \beta B)= \alpha (L_1, L_2, L_3)\cdot A
 + \beta (L_1, L_2, L_3) \cdot B\;.
 $$
 \item For $L_i \in \C^{m_i \times n_i}, M_i \in \C^{l_i \times m_i}$, 
 $i=1, \cdots, k$, 
 $$
 (M_1, M_2, M_3) \cdot [(L_1, L_2, L_3) \cdot A]=
 (M_1 L_1, M_2 L_2, M_3 L_3)\cdot A\;.
 $$
\item For any $M_i, N_i \in \C^{m_i \times n_i}$, $\alpha, \beta \in \C$,
$$
(\alpha M_1 + \beta N_1, L_2, L_3) \cdot A=
\alpha ( M_1, L_2, L_3) \cdot A + \beta ( N_1, L_2, L_3) \cdot A\;,
$$
and similarly for the other two slots. 

\end{enumerate}

\vspace{3mm}
Before going to their applications, we work with cubic hypermatrices 
of general `size' $n$ for which we have the following result

\begin{prop} \label{discrim}
  Let $A\in\R^{n\times n\times n},$ let $A'$ be obtained from $A$ by
  permuting the three factors in the tensor product, and let
  $(L,M,N)\in GL_n(\R)^{\times 3}$. Then $\Delta(A')=\Delta(A)$ and
  $\Delta((L,M,N)\cdot A)=\det(L)^n\det(M)^n\det(N)^n \Delta(A).$
\end{prop}

\noindent This is a generalization of Proposition 5.6 in \cite{DL}, and the proof
is similar. Here $\Delta$ is the discriminant defined 
right after the proof of proposition \ref{propo 5} below and 
$G^{\times k}$ is the product $\underbrace{G \times \cdots \times G}_k$.

\vspace{3mm}
\noindent {\bf Remark.}
The hyperdeterminant as defined in expression \eqref{hyperdet}
is not the same as the discriminant of a tensor. 
The notion used in Corollary 1.5 of \cite{GKZ} is the 
discriminant $\Delta (A)$. In fact, both $\mathcal{D}$et and $\Delta$ are invariant
under matrix operations. More precisely, they are invariant under the 
action of $SL_n^{\times k}$. However,
the polynomial $\Delta$ is in general much more complicated than 
$\mathcal{D}$et. For example, if one considers $2\times 2 \times 2 \times 2$
hypermatrices, then $\mathcal{D}et$ is a polynomial of degree
2 while $\Delta$ is a polynomial of degree 24.

\vspace{3mm}
\noindent {\bf Symmetric tensors and hypermatrices.}
A 3-dimensional cubic hypermatrix $A=[a_{ijk}]\in \C^{n \times n \times n}$ 
is symmetric if 
$a_{ i_{\sigma (1)} i_{\sigma (2)} i_{\sigma (3)} }=a_{i_1 i_2 i_3}$, 
with $i_1$, $i_2$, $i_3\in \{1, \cdots, n\}$ for all permutations 
$\sigma$ of the symmetric group $\Sigma_3$. Explicitly this is 
\(
a_{ijk}=a_{ikj}=a_{jik}=a_{jki}=a_{kij}=a_{kji}, ~~~\forall~ i, j, k \in \{1, \cdots, n\}\;.
\)
An order three tensor  $A \in \C^n \otimes \C^n \otimes \C^n$ is symmetric
if $\sigma (A)=A$ for all permutations $\sigma \in \Sigma_3$, where the group
action is given by
\(
\sigma ( x_{i_1} \otimes x_{i_2} \otimes x_{i_3})= x_{i_{\sigma (1)}} 
\otimes x_{i_{\sigma (2)}} 
\otimes x_{i_{\sigma (3)}}\;. 
\)
Given a basis $\{ e_{1} \cdots e_{n}\}$ of 
$\C^n$ then a basis of the set $S^3 (\C^n)$ of symmetric 3-tensors in 
$\C^n$ is given by 
\(
\left\{
\frac{1}{3!} \sum_{\sigma \in \Sigma_3} e_{i_{\sigma (1)}} \otimes 
e_{i_{\sigma (2)}} \otimes e_{i_{\sigma (3)}} ~|~ 1 \leq i_1 \leq i_2 \leq i_3 \leq n
\right\}\;,
\)
whose (complex) dimension is 
\(
\dim_{\C} S^3 (\C^n)= 
\left(
\begin{array}{c}
n+2
\\
3
\end{array}
\right)=\frac{1}{6}n (n+1)(n+2)
\;.
\)
This corresponds to the number of partitions of 3 into a sum of $n$ 
nonnegative integers, so that for $n=1, 2, 3$ this is 
$1, 4$, and $10$, respectively.  

\vspace{3mm}
There is a bijective correspondence between symmetric tensors and homogeneous 
polynomials of degree three in n variables 
\(
S^3(\C^n) \cong \C [x_1, \cdots, x_n]_3\;.
\)
For $n=1$ this is $\C[x]_3$ which is of the form $x^3$. For $n=2$ this is 
$\C [x, y]_3$ which is formed of the four monomials $x^3$, $x^2 y$, $xy^2$, $y^3$. 
For $n=3$ this is formed of the ten monomials $y z^2$, $xz^2$, $y^2 z$,
$x y^2$, $x^2z$, $x^2 y$, $x y z$, $x^3$, $y^3$, $z^3$.

\vspace{3mm}
\noindent {\bf Direct sum.} The direct sum of two order-3
  tensors$/$hypermatrices $A \in \C^{l_1 \times m_1 \times n_1}$ and
  $B \in \C^{l_2 \times m_2 \times n_2}$ is a ``block tensor" with $A$
  in the $(1,1,1)$-block and $B$ in the $(2,2,2)$-block \( A \oplus B
  = \left[
\begin{array}{cc|cc}
$A$ & 0 & 0 &0\\
0& 0& 0& $B$
\end{array}
\right]
\in \C^{(l_1 + l_2) \times (m_1 + m_2) \times (n_1 + n_2)}\;.
\end{equation}

In terms of vector spaces, if $U_i$, $V_i$ , $W_i$ are vector spaces
such that $W_i=U_i \oplus V_i$ for $i=1, \cdots, k$, then the tensors
$A \in U_1 \otimes U_2 \otimes U_3$ $B \in V_1 \otimes V_2 \otimes
V_3$ have direct sum $A \oplus B \in W_1 \otimes W_2 \otimes W_3$.

\vspace{3mm}
\noindent {\bf Tensor rank.} Such a notion goes back as far as
reference \cite{Hi}. A tensor has a tensor rank $r$ if it can be
  written as a sum of $r$ decomposable tensors, but no fewer \(
  {\rm rank}_{\otimes} (A):= {\rm min} \left\{ r~|~ A=\sum_{i=1}^r u_i
    \otimes v_i \otimes \cdots \otimes z_i \right\}\;.  \) A nonzero
  decomposable tensor has tensor rank 1.

\vspace{3mm} 
We have the following general result on tensor rank.
\footnote{The referee informed us that he thinks that this might have appeared before, 
perhaps in \cite{Hi}.}

\begin{prop} Let $A\in\mathbb{C}^{n_{1}\times\cdots\times n_{k}}$. Then 
\[
\mathrm{rank}\,\left(A\right)\leq\dfrac{{\displaystyle
    \prod_{i=1}^{k}n_{i}}}{{\displaystyle \max_{i}}\left\{
    n_{i}\right\} }.\] 
    \label{propo 5}
    \end{prop}

\proof
  $k=2$ is obvious. We proceed by induction on $k-1$. Define $l$ by
  $n_{l}={\displaystyle \min_{i}}\left\{ n_{i}\right\} $. Without loss
  of generality, $l=1$. Each slice
  $A_{t}=\left(a_{\left(t-1\right)i_{2}\cdots i_{k}}\right)$ for
  $t=1,\ldots,n_{1}$ is a $(k-1)$--dimensional hypermatrix. By our
  induction hypothesis, each of these slices has rank at most 
  \(
  \dfrac{{\displaystyle \prod_{i=2}^{k}n_{i}}}{{\displaystyle
      \max_{i}}\left\{ n_{i}\right\} }\;.
      \)
       Write \(
  A_{t}={\displaystyle
    \sum_{m=1}^{r}[v_{t_{1}}\otimes\cdots\otimes
      v_{t\left(k-1\right)}}]_m\;,
      \) 
      and express
  $A=\vec{e_{1}}\otimes A_{1}+\cdots+\vec{e}_{n_{l}}\otimes
  A_{n_{l}}$.  Expanding the sum, the result follows.
\endofproof

\vspace{3mm}
The discriminant can be defined for homogeneous 
forms in $k+1$ variables of degree $d$ as follows \cite{GKZ}. 
The discriminant is an irreducible polynomial $\Delta (f)$ 
in the coefficients of a form $f=f(x_0, x_1, \cdots, x_k)$ which 
vanishes if and only if all the partial derivatives 
$\partial f/\partial x_0$, $\partial f/ \partial x_1$, $\cdots$, $\partial f/\partial x_k$
have a common zero in $\C^{k+1}-\{0\}$. Note that $\Delta (f)$ 
depends on the degree $d$. The requirement that the 
polynomial $\Delta (f)$ be irreducible over $\Z$, i.e. that it has relatively
prime integer coefficients, makes it  
defined uniquely up to a sign. The importance of the discriminant is that
is that it vanishes whenever $f$ has multiple roots. This is familiar from the
low degree cases, namely the quadratic and cubic polynomials. 
For a tensor $A$, the discriminant $\Delta (A)$
is the hyperdeterminant. We have the following generalization of 
Proposition 5.9 in \cite{DL}

\begin{prop}
Let $A \in \R^{n \times n \times n}$. If $\Delta(A) > 0$ implies that
$\det(\sum^{n}_{i=1} \lambda_{i} A_{i})$ has $n$ distinct real sets of roots,
then ${\rm rank}(A) \leq n(n-1)$.
\end{prop}

\proof
  By hypothesis, we have $n$ distinct real sets of roots for
  $\det(\sum^{n}_{i=1} \lambda_{i} A_{i})$, for $i=1,\ldots ,n$,
  $\lambda_{i1}, \ldots, \lambda_{in}$.  Then we can transform
  $\left[\begin{array}{c|c|c} A_{1} & \cdots & A_{n}
\end{array}\right]$
by slab operations into
$\left[\begin{array}{c|c|c}
B_{1} & \cdots & B_{n}
\end{array}\right] = B$,
where $B_{i} = \sum^{n}_{j=1} \lambda_{ij} A_{j}$. By construction,
$\det(B_{i}) = 0$, so $B_{i}$ is of non-maximal rank, such that
$B_{i} = \sum^{n-1}_{s=1} f_{is} \otimes g_{is}$. Taking the tensor product
$e_{i} \otimes B_{i}$ and summing over all $i$ gives an expression of $n(n-1)$
rank-1 hypermatrices for $B$. Since rank is invariant under Gaussian processes,
${\rm rank}(A) = {\rm rank}(B) \leq n(n-1)$.
\endofproof

\subsection{The action of the general linear group on wedge products}
The natural action of $GL(V)$ on $V$ extends canonically to the 
exterior powers of $V$. For completeness we review this briefly, following
\cite{VP}.
 The elements of $\wedge^m (V)$ are called $m$-vectors of polyvectors
of degree $m$. Polyvectors which can be written in the form $u_{i_1} \wedge \cdots \wedge 
u_{i_m}$ for some vectors $u_1, \cdots, u_m$ are called {\it decomposable 
polyvectors}. 
 On decomposable polyvectors, multiplication is defined by the formula
\(
(u_1 \wedge \cdots \wedge u_k, x_1 \wedge \cdots \wedge x_m) \mapsto
u_1 \wedge \cdots \wedge u_k \wedge x_1 \wedge \cdots \wedge x_m\;. 
\)
In particular, the degree of the product  equals the sum of the degrees of its
factors. 

\vspace{3mm}
Elements of $\wedge^m (V^*)$ are exterior (differential) $m$-forms. The exterior power 
$\wedge^m (V^*)$ can be identified with $\wedge^m (V)^*$ by means of the
canonical pairing. On the decomposable 
polyvectors and decomposable exterior forms, this pairing is given by 
\(
(v_1 \wedge \cdots \wedge v_m, u_1 \wedge \cdots \wedge u_m) \mapsto 
\det (v_i(u_j))\;, ~~~ v_1, \cdots, v_m \in V^*, ~~ u_1, \cdots, u_m \in V\;.
\)
The pairing uniquely extends to the pairing between exterior algebras 
$\wedge (V)$ and $\wedge (V^*)$ the image of which on polyvectors and 
exterior forms of distinct degrees equals zero.
 Choosing a basis $e_1, \cdots, e_m \in V$, we can identify the automorphism
group $GL(V)$ of the module $V$ with $GL(n, \mathcal{R})$. For every $m$, the group
$GL(n, \mathcal{R})$ acts naturally on $\wedge^m (V)$. The action of 
$g \in GL(n, \mathcal{R})$ on decomposable $m$-vectors is given by 
\(
\wedge^m (g) (v_1 \wedge \cdots \wedge v_m)= gv_1 \wedge \cdots \wedge gv_m\;,
~~~~\forall~ v_1, \cdots, v_m \in V.
\)

\vspace{3mm}
The Binet-Cauchy theorem asserts that the map
$
\wedge^m : GL(n , \mathcal{R}) \longrightarrow 
GL(C_n^m, \mathcal{R})$, 
$
g \mapsto  \wedge^m (g)
$,
where $C_n^m$ are the binomial coefficients, is in fact a homomorphism
\(
\wedge^m(hg)= \wedge^m (h) \wedge^m (g)\;.
\)
Thus the map $g \mapsto \wedge^m (g)$ is a degree-$C_n^m$ 
representation of the group $GL(n, \mathcal{R})$ \cite{VP}. It is called the 
$m$-vector representation of the $m$th {\it fundamental representation}. 

\vspace{3mm}
As in section \ref{rep groups},
consider the following subgroup of the group $G$ corresponding to the 
Lie algebra $\frak{g}$
\(
G^{\theta}=\left\{ g \in G~|~ \theta (g)=g\right\}\;.
\)
Let $G_0 \subset G^{\theta}$ be the group corresponding to the 
subalgebra $\frak{g}_0$. 
From the property $[\frak{g}_0~,~ \frak{g}_k] \subset \frak{g}_k$
it follows that the adjoint representation of the group $G$ induces, by restriction,
a linear representation $\rho_k$ of $G_0$ in $\frak{g}_k$ (for any $k$)
\cite{V1}.

\subsection{Ranks and orbits of $3$-vectors: Admissible dimensions for $V$}



 A generic $3$-tensor is an element of $\otimes^3 (\C^n)$
with open $GL(n, \C)$ orbit. Similarly for symmetric and 
antisymmetric powers.
The isotropy group of a tensor consists of all group elements leaving the
tensor invariant, 
\(
G_T := \left\{ A \in GL(n, \C) ~|~ T = AT \right\}\;.
\)
The dimension of this space is $n^2- \dim \left( \otimes^3 (\C^n)\right)$.
Similarly for the antisymmetric and symmetric powers, in which cases
the tensor power $\otimes^3$ is replaced by either $\wedge^3$ or $S^3$, 
respectively.


\vspace{3mm}
 Consider the orbits of the group $\wedge^m (GL(n, \mathcal{R}))$ 
acting on $\wedge^m (V)$. 
 For bivectors the situation is very simple as every bivector is equivalent to 
one of the bivectors $e_1 \wedge e_2 + \cdots + e_{2r-1} \wedge e_{2r}$,
$1 \leq 2r \leq n$, under the action of $\wedge^m(GL(n, \mathcal{R})$.
We are interested mainly in the cases when $\mathcal{R}$ is $\R$ or $\C$.

\vspace{3mm}
The rank of an orbit of an $m$-vector in an $n$-dimensional vector space can take 
only the
values (see e.g. \cite{23})
\(
0, m, m+2, \cdots, n,
\)
so that for trivectors the only possible ranks are: 
$0, 3, 5, \cdots, n$. 

\vspace{3mm}
\noindent {\bf The complex case:} The orbits of a given rank are
known and are described as follows. 

\noindent {\it Rank 0:} only $0$ is possible.

\noindent  {\it Rank 3:} only $e_1 \wedge e_2 \wedge e_3$.

\noindent  {\it Rank 5:} any such trivector is equivalent to 
$
e_1 \wedge e_2 \wedge e_3 + e_1 \wedge e_4 \wedge e_5$.

\noindent  {\it Rank 6:} There are two orbits of complex trivectors of rank 6, with representatives
(Reichel 1907)
\bea
e_1 \wedge e_2 \wedge e_3 + e_4 \wedge e_5 \wedge e_6\;,
\nonumber
e_1 \wedge e_2 \wedge e_4 + e_1 \wedge e_3 \wedge e_5 + e_2 \wedge e_3 \wedge e_6\;.
\eea

\noindent  {\it Rank 7:} The complex trivectors of rank 7 have 5 orbits 
(J. A. Shcouten 1931). 

\noindent  {\it Rank 8:} The complex trivectors of rank 8 have 13 orbits (Gurevich 1935).

\noindent  {\it Rank 9:} In this case two new interesting features occur \cite{VE}:
First, there are infinitely many orbits.
Second, here close connections to Lie algebras start to become apparent. 
This uses the exceptional embedding $A_8 \subseteq E_8$ and the graded Lie algebra decomposition
\bea
\frak{e}_8=\frak{g}&=&\frak{g}_{-1} \oplus \frak{g}_0 \oplus \frak{g}_1
\nonumber\\
&=& 
\wedge^3V^* \oplus \frak{sl}(V) \oplus \wedge^3 V\;,
\eea
where $V=\C^9$. 
There is a nontrivial homomorphism of $SL(9, \C)$ 
into the adjoint group of $\frak{g}$ whose kernel is the central subgroup
of order 3.
This action of $SL(9, \C)$ on $\frak{g}$ preserves the grading. 
Restricting the action to $\frak{g}_1$ gives the desired action of 
$SL(9, \C)$ on $\wedge^3 (\C^9)$. 
The general method of Vinberg \cite{V1} \cite{V2} can be applied to classify the orbits of 
$SL(9, \C)$ in $\frak{g}_1$.


\vspace{3mm}
\noindent {\bf The real case.} 
 $\wedge^3 (\R^n)$ is a real subspace of the complexification
$\wedge^3(\C^n)$.
For $x \in \wedge^3(\R^n)$ the real orbit $GL(n, \R)\cdot x$ is contained 
in the complex orbit $GL(n, \C)\cdot x$. This orbit is called a {\it real form} of the complex 
orbit containing it. 
Every complex orbit has only finitely many real forms
\cite{BH}.
The problem of classifying the orbits of $GL(n, \R)$ in $\wedge^3(\R^n)$ thus reduces 
to the problem of classifying the real forms of the orbits of $GL(n, \C)$ in $\wedge^3(\C^n)$. 
The orbits of a given rank are
known and are described as follows. 

\noindent {\it Rank $\leq 5$:}
the classification is trivial. 

\noindent  {\it Rank 6:}
The classification is obtained by Gurevich in the 1930's, then by 
\cite{Ca} and \cite{Re}.

\noindent  {\it Rank 7:}
Given in  \cite{We} and \cite{Re}.

\noindent  {\it Rank 8:} All real forms of the 23 orbits of $GL(8, \C)$ in 
$\wedge^3(\C^8)$ are enumerated in \cite{Dj}. 

\vspace{3mm}
From the above classic results, we state the following
\begin{prop}
In representing three-junctions by finite-dimensional exceptional
Lie algebras (or groups) according to the graded Lie algebra decomposition, 
the highest dimension for the corresponding vector space $V$ is 9.   
\end{prop}




\vspace{3mm}
\noindent {\bf Interpreting the orbits.}
The classification of the orbits of the action of the general linear group
on $\frak{g}_1$, which generically is either a symmetric, tensor, or antisymmetric
power of some vector space, involves considering a three-dimensional 
subalgebra \cite{VE} \cite{Dj}. The elements $h \in \frak{g}_0$,
$x \in \frak{g}_1$, and $y \in \frak{g}_{-1}$ form a graded $\frak{sl}_2$-triple
$(x, h, y)$ with
\(
[h, x]=2x\;, \qquad \qquad [h, y]=-2y\;, \qquad \qquad [x,y]=h\neq 0\;.
\)
The vector space will decompose as $V=V_1 \oplus \cdots \oplus V_k$,
with $k$ depending on the algebra and its grading.
The centralizer $Z(h)$ will be of the form 
$SL_{m_1} \times \cdots \times SL_{m_k}$, with 
$m_1 + \cdots + m_k={\rm dim}(V)$, where
$SL_{m_i}=SL(V_i)$, $1 \leq i \leq k$. 

\vspace{3mm}
We interpret $Z(h)$ as the breaking of the original symmetry
$SL(V)$ into the corresponding product pieces. What this means
dynamically is that we are moving apart the stack of branes 
into a final $m_k$ sets of thinner stacks. For example, start with
a decomposition $V=V_1 \oplus V_2$ so that $SL(V)$ breaks into
$SL(V_1) \times SL(V_2)$. This decomposition means, for the ends of 
the junction, that  the original stack of branes corresponding to $V$
will break into two separate stacks, formed (for the case of D-branes)
of ${\rm dim}(V_1)$
and ${\rm dim}(V_2)$ overlapping branes, respectively.   
The tensor product $V^{\otimes 3}$ will
involve pieces $V_1^{\otimes 3}$, corresponding to a junction
joining the ${\rm dim}(V_1)$ stacks together, and $V_2^{\otimes 3}$,
corresponding to a junction joining the other stacks, i.e. the
one with ${\rm dim}(V_2)$ components. In addition, there are
`mixed junctions', i.e. ones which connect $i$ stacks of first type to 
$(3-i)$ stacks of the second type, with $i=1,2$. The extension to the
general case is straightforward. Thus we see that one can have more
junction configurations, starting with basic ones which correspond to 
$\frak{g}_0$.

\begin{prop}
The $SL(V)$-action on $\frak{g}_1$ 
leads, via breaking of symmetry, to admissible junction configurations 
according to the corresponding orbits. 
\end{prop}

\subsection{The traces and invariants}
\label{tr}

 We have seen that single strings can be represented simply as matrices
  ${\l^i}_j$.  When joining multiple strings together the indices 
  indices which correspond to adjoined ends are contracted.  
  \(
  {(\l^{(1)})^{i_1}}_{i_2}{(\l^{(2)})^{i_2}}_{i_3}\cdots{(\l^{(m)})^{i_m}}_{i_1}
  = \mathrm{tr}(\l^{(1)}\l^{(2)}\cdots\l^{(m)}).  \)
 Similarly, as we saw in section \ref{case junction}, 
 we use a cubic hypermatrix $\l_{ijk}$ to represent
  three-pronged junctions.  An analogous expression for trace is then
  \( \sum_{i,j,k=1}^n \l^{(1)}_{ijk}\l^{(2)}_{jki}\l^{(3)}_{kij}. \)
  We set up the indices cyclically so as to satisfy the conservation
  of charge condition for each endpoint.

\vspace{3mm}
Notice that if we model the vertex by $\wedge^3 V$, our
  generalized expression for trace is trivial, for by renaming the
  indices and applying antisymmetry we get
  \bea  
  \sum_{i,j,k=1}^n
    \l^{(1)}_{ijk}\l^{(2)}_{jki}\l^{(3)}_{kij} &= &\sum_{i,j,k=1}^n
    \l^{(1)}_{ikj}\l^{(2)}_{kji}\l^{(3)}_{jik}\\
    \nonumber\
    &=& \sum_{i,j,k=1}^n
    (-\l^{(1)}_{ijk})(-\l^{(2)}_{jki})(-\l^{(3)}_{kij})\;,
  \eea
  whence it is zero.  


\vspace{3mm}
We would like to consider the question of admissible traces more
systematically, using invariant theory, as follows. Any $\lambda \in \frak{g}_1$ 
admits a {\it Jordan decomposition} $\lambda^{\rm ss} + \lambda^{\rm nil}$
into a semisimple part $\lambda^{\rm ss}$ and a nilpotent part
$\lambda^{\rm nil}$. The former type is characterized by the property that 
their orbit  is closed, while the latter elements are determined by the
property that the closure of their orbit contains $0$. The {\it Cartan subspace}
is a maximal subspace $\frak{c}\subset \frak{g}_1$ consisting of commuting
semisimple elements
\(
\frak{c}=\left\{ 
\lambda^{\rm ss}_1, \lambda^{\rm ss}_2 \in \frak{g}_1 ~|~
\left[ \lambda^{\rm ss}_1~,~\lambda^{\rm ss}_2 \right]=0
\right\}\;.
\)
Thus a basis for $G=D_4$ will be composed of symmetric polynomials
in three variables $x_1$, $x_2$, $x_3$, 
that of $G=E_6$ will be composed of triple tensor products of the 
vectors $e_1$, $e_2$, $e_3$ that form a basis for $\C^3$, while that of 
$G=E_8$ will be wedge products of the vectors $e_1, \cdots e_9$ that 
for a basis for $\C^9$. 

\vspace{3mm}
The {\it Weyl group} is the group $W$ of linear transformations of $\frak{c}$ 
generated by elements of $G_0$ the adjoint action of which leaves
$\frak{c}$ invariant. 

\vspace{3mm}
The algebra $\C [\frak{g}_1]^{G_0}$ of $G_0$-invariant polynomials
in $\frak{g}_1$ is free and is isomorphic to the algebra 
$\C [\frak{c}]^W$ of Weyl-invariant polynomials in the Cartan subspace
\cite{V2}. We will consider specific examples in the next section.

\begin{figure}
\centering
\includegraphics[scale=.8]{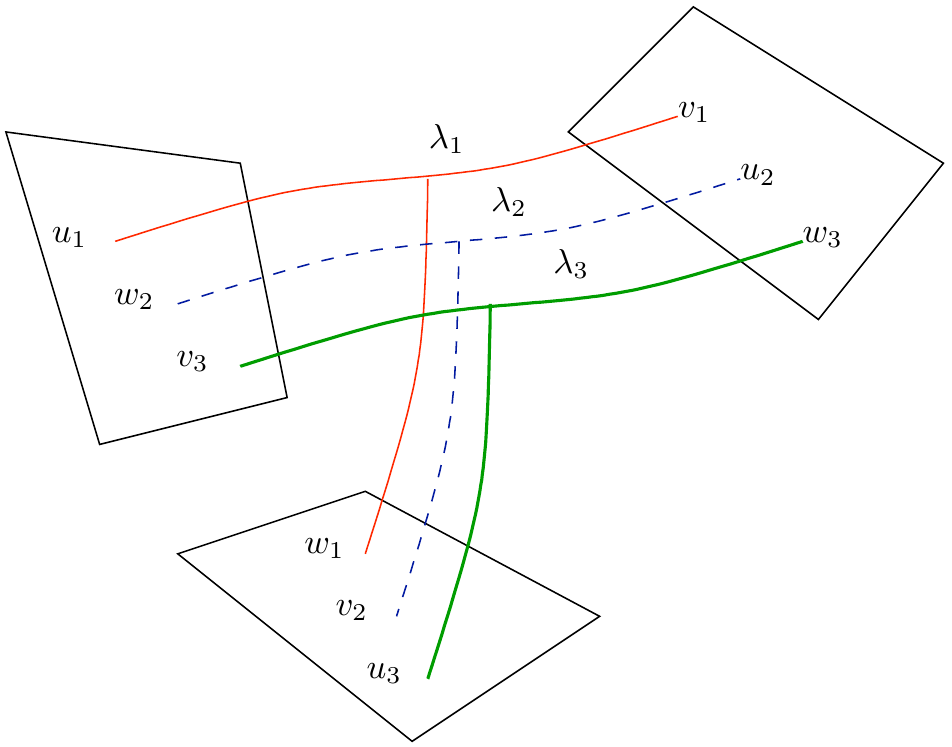}
\caption{Contraction of three junctions.}
\end{figure}

\begin{figure}
\centering
\includegraphics[scale=.8]{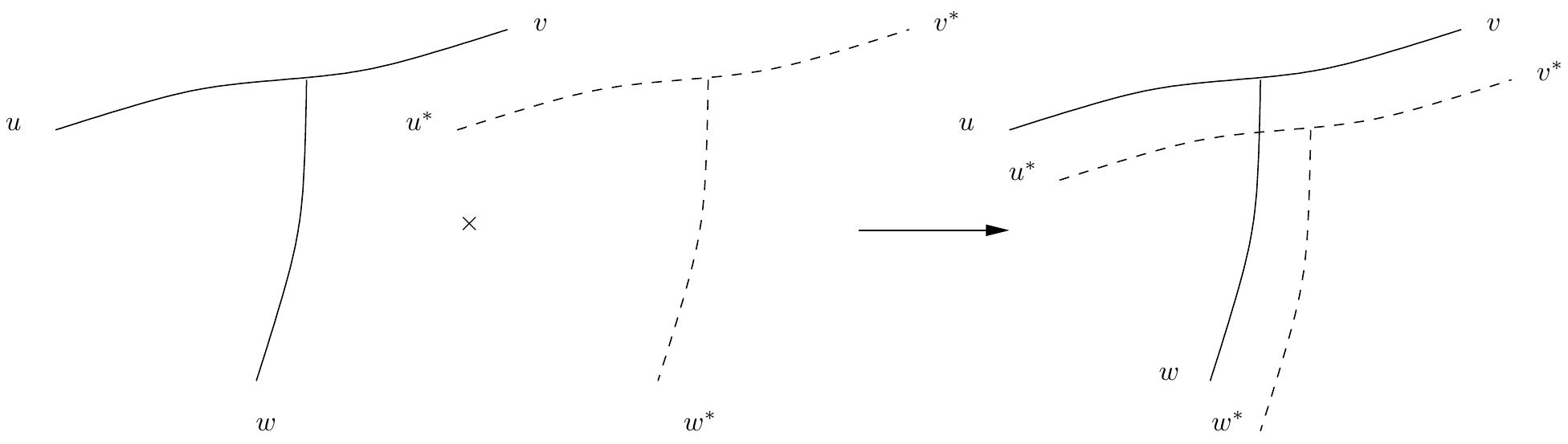}
\caption{Commutator of a junction and its dual.}
\end{figure}

\section{The Main Examples}

\vspace{3mm}
In case the vector spaces $V$ are of dimensions less than or equal to 9, we have
\begin{prop}
String and membrane three-junctions provide (and hence can be described by)
representations of $GL(9,\C)$ (and hence of its
subgroups by restriction) on $\wedge^3 V$. (Similarly for the 
compact subgroups $SU(9)$ and their subgroups when 
requiring that norms of states be preserved).
\end{prop}

\vspace{3mm}
Given proposition \ref{prop case} we see that $\frak{sl}(V)$ arises as the
$\frak{g}_0$ factor in the graded decomposition of $\frak{e}_8$,
$\frak{e}_6$ and $\frak{d}_4$. Thus, it is natural to consider these
Lie algebras.
We summarize the main result of the examples in the following 
four sections, i.e. sections \ref{e8 ex}, \ref{e6 ex}, \ref{d4 ex},
and \ref{f4 ex} as
\begin{theorem}
String and membrane 3-junctions allow for $\frak{d}_4$, $\frak{e}_8$, 
$\frak{f}_4$, or $\frak{e}_6$ 
 symmetries, depending on whether we take hypermatrix factors
for the junctions to be symmetric, antisymmetric, symmetric on two indices, 
or of no symmetries, respectively. 
\end{theorem}

\subsection{Representations of the Lie algebra $\frak{g}_0$
on $\wedge^3 V$: The $E_8$ example}
\label{e8 ex}

 Let $V$ be a vector space and $V^*$ the dual vector space to $V$.
Consider $\wedge^3 V$, the third exterior power of $V$. This can be 
identified with $V^{\otimes 3}=V \otimes V \otimes V$,
the space of 3rd tensor power of $V$, so that for any $v_1, v_2, v_3 \in V$,
\(
v_1 \wedge v_2 \wedge v_3=\sum_{\rm perm.} {\rm sgn} (i_1, i_2, i_3)
v_{i_1} \otimes v_{i_2} \otimes v_{i_3}\;. 
\)
Form the third exterior power $\wedge^3 V^*$ of $V^*$. There is a duality
between $\wedge^3 V$ and $\wedge^3 V^*$ (using Einstein's summation convention
henceforth):
\(
\langle \lambda, \lambda^* \rangle = 
\frac{1}{3!} \lambda^{i_1 i_2 i_3} \lambda^*_{i_1 i_2 i_3}\;,~~~~~~
\lambda \in \wedge^3 V, ~~\lambda^* \in \wedge^3V^*\;.
\)
 Similarly, if $\epsilon$ is a nonzero element of the space $\wedge^9 V$, then
  $\epsilon^*$ will denote the element of the space $\wedge^9 V^*$ that satisfies 
  $\langle \epsilon, \epsilon^* \rangle=1$.   
 Let $L(V)=V\otimes V^*$ be the space of linear transformations of $V$ and 
\(
L_0(V)=\left\{ S \in L(V)~|~ {\rm Tr} (S)=0 \right\}\;.
\)
These form the algebras $\frak{gl}(V)$ and $\frak{sl}(V)$, respectively.

  
  \vspace{3mm}
 To each of the graded Lie algebra decompositions, we associate Lie
  commutators. When writing equations 
  explicitly we will use component notation. 
  For $\mathfrak{e}_{8}$, with $X, Y \in
  \mathfrak{g}_{0}$, $\l, \l_{1}, \l_{2} \in \mathfrak{g}_{1}$, and
  $\l^{*}, \l^{*(1)}, \l^{*(2)} \in \mathfrak{g}_{-1}$, the commutation 
  relations, which result from breaking the original Lie bracket on $\frak{g}$
  into components corresponding to the grading, are \cite{VE}
  \bea
  \left[X~,~Y\right]^{i}_{j} &=& X^{i}_{s}Y^{s}_{j} -
    Y^{i}_{s}X^{s}_{j} \qquad \qquad \qquad \qquad \quad
    \quad \hspace{2mm}
     \subset \mathfrak{g}_{0}
    \\
    \left[X~,~\l \right]^{ijk} &=& X^{i}_{s}\l^{sjk} + X^{j}_{s}\l^{isk} +
    X^{k}_{s}\l^{ijs} \qquad \qquad 
    \quad
    \subset \mathfrak{g}_{1}
    \label{ac}
    \\
    \left[X~,~\l^{*}\right]_{ijk} &=& -\l^{*}_{sjk}X^{s}_{i} -\l^{*}_{isk}X^{s}_{j}
    -\l^{*}_{ijs}X^{s}_{k} \qquad \qquad \hspace{1mm} \subset \mathfrak{g}_{-1}
    \\
    \left[\l_{1}~,~\l_{2}\right]_{ijk} &=& \frac{1}{(3!)^2}
    \epsilon^{*}_{p_{1}q_{1}r_{1}p_{2}q_{2}r_{2}ijk}
    \l^{p_{1}q_{1}r_{1}}_{1} \l^{p_{2}q_{2}r_{2}}_{2} 
    \qquad 
    \subset
    \mathfrak{g}_{-1}
    \label{cc}
    \\
    \left[\l^{*(1)}~,~\l^{*(2)}\right]^{ijk} &=& -\frac{1}{(3!)^2}
    \epsilon^{p_{1}q_{1}r_{1}p_{2}q_{2}r_{2}ijk}
    \l^{*(1)}_{p_{1}q_{1}r_{1}} \l^{*(2)}_{p_{2}q_{2}r_{2}}
    \quad 
    \hspace{1.3mm}
     \subset
    \mathfrak{g}_{1}
    \label{c*c*}
    \\
    \left[ \l~,~\l^{*} \right]^{i}_{j} &=& \frac{1}{2} \l^{pqi} \l^{*}_{pqj} - \frac{1}{18}
    \l^{pqr} \l^{*}_{pqr} \delta^{i}_{j} \qquad \qquad 
    \quad \hspace{1mm}
    \subset \mathfrak{g}_{0}\;,
  \label{g1g-1}
  \eea
where $\epsilon^{p_{1}q_{1}r_{1}p_{2}q_{2}r_{2}ijk}$
and $\epsilon^{*}_{p_{1}q_{1}r_{1}p_{2}q_{2}r_{2}ijk}$ are the components
of $\epsilon$ and $\epsilon^*$, respectively. 
  Alternatively, if $\l = x \wedge y \wedge z$ and $\l^{*} = f \wedge g
  \wedge h$, the last commutator can be written
  \begin{gather*} [x \wedge y \wedge z, f \wedge g \wedge h] = -
    \begin{vmatrix}
      f(x) & f(y) & f(z) & f \\
      g(x) & g(y) & g(z) & g \\
      h(x) & h(y) & h(z) & h \\
      x & y & z & \frac{1}{3}I
    \end{vmatrix} .
  \end{gather*}

\vspace{3mm}
\noindent {\bf Remark.}
%
An important question to ask is whether the factor $\frak{g}_1=\wedge^3V$
forms an algebra by itself. The answer is no, as hinted earlier. However, 
while this is not the case we see
from the commutation relation (\ref{cc}) that two factors in $\wedge^3 V$ 
close into an element of $\wedge^3 V^*$, which is a degree three element
but for the dual vector space. Thus, this process does produce 
the desired form provided that we also introduce the operation of 
dualization for the vector spaces. Similarly, for starting with the dual vector space
the commutations relation (\ref{c*c*}) of two dual degree three forms gives a
degree three form of the original vector space. We will see another model for
the multiplication of two three-forms in section \ref{duality symmetry}.

\begin{figure}
\centering
\includegraphics[scale=.8]{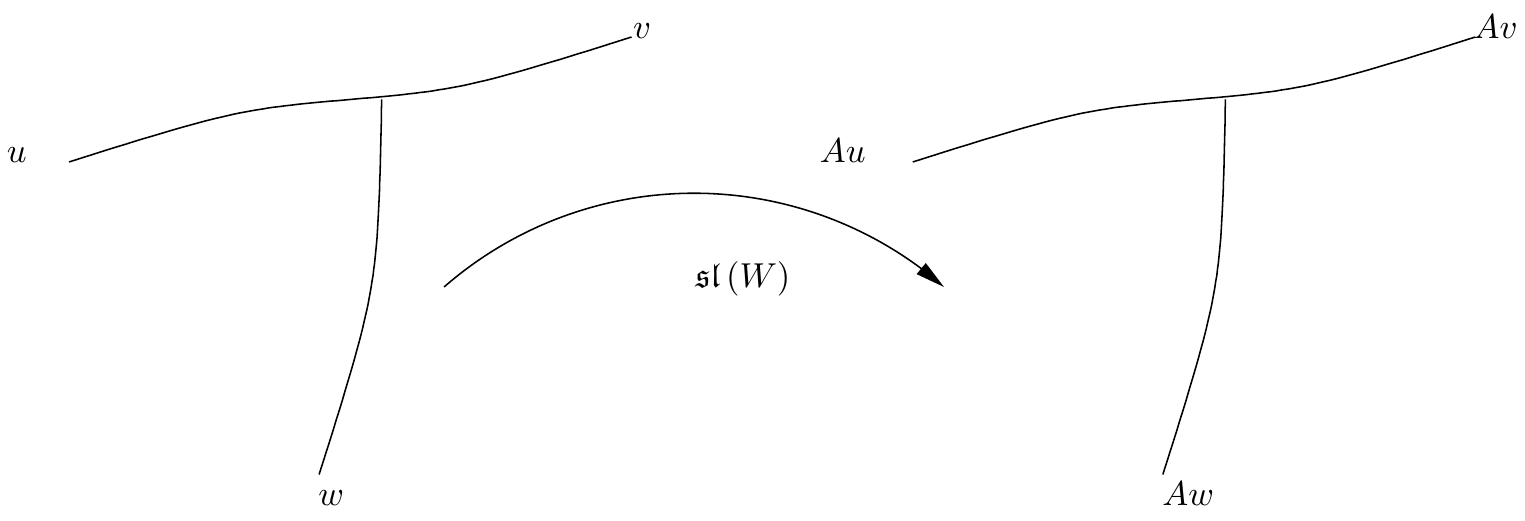}
\caption{Action of the algebra on the string.}
\end{figure}

\vspace{3mm}
\noindent {\bf Which traces can occur?}
As mentioned in section \ref{tr} we will consider traces using invariant theory. 
The free generators of the invariant algebra of the action of 
$G_0=SL(V)$ on $\frak{g}_1=\wedge^3 V$, 
${\rm dim} (V)=9$, have degrees \cite{V2}
12, 18, 24, 30. They can be constructed as follows \cite{Eg} \cite{Ka}.
Consider the linear transformation
\(
\mathcal{L}_\otimes (\lambda): V \otimes V \otimes V 
\longrightarrow V \otimes V \otimes V\;.
\label{lin vvv}
\)
Since $\wedge^3 V \subset V \otimes V \otimes V$ 
(cf. equation (\ref{triple}))
then the restriction of the above linear transformation 
to $\wedge^3 V$ is
\(
\mathcal{L}_\wedge (\lambda) : \wedge^3 V \longrightarrow 
\wedge^3 V\;,
\)
which is the cube of the action of $\lambda
\in \wedge^3 V$ on $\wedge^3 V^*$, given 
in (\ref{g1g-1}).
  Starting with 
$\lambda \in \wedge^3 V$, 
the tensor  defining the linear transformation
(\ref{lin vvv}) is of type $(3,3)$ and given by
\(
\left(C(\lambda) \right)_{i_1 i_2 i_3}^{mnp}
=\epsilon_{l_1 l_2 l_3 i_1 i_2 i_3 j k_1 k_2}
\lambda^{l_1 l_2 l_3} \left( 
\lambda^{mk_1 k_2} \lambda^{jnp} +
\lambda^{nk_1 k_2} \lambda^{jmp} +
\lambda^{pk_1 k_2} \lambda^{mnj} 
\right)\;.
\) 
This tensor is skew-symmetric in both superscripts and
subscripts.

\vspace{3mm}
One can take for the generators $P_1, \cdots, P_r$
of the algebra of invariants of the adjoint representation of 
the algebra $\frak{e}_8$ to be the trace of the 
$k$th power of the action of an element of $\frak{e}_8$ 
on $\frak{e}_8$. In our case $k=3$ and the trace of 
$\mathcal{L}_\otimes$ coincides with that of $\mathcal{L}_\wedge$. 

\vspace{3mm}
By general results of \cite{Ka}, the restriction of the 
algebra $\C [\frak{e}_8]^{E_8}$ on 
$\wedge^3 V$ coincides with 
$\C [\wedge^3V]^{SL(V)}$, and the degrees $n$ of the 
free generators are $12, 18, 24$ and $30$. 
The explicit form of the generators is given in 
\cite{Ka} \cite{Eg} as
\(
f_{3n} (\lambda)={\rm tr} {\mathcal{L}}_\otimes (\lambda)^n\;, \qquad n=4, 6, 8, 10.
\label{f3n}
\)

\vspace{3mm}
Because of the isomorphism $\C [\frak{g}_1]^{G_0} \cong \C [\frak{c}]^W$, we
can also look at the invariants using Weyl invariance instead. 
For
$\frak{e}_8$, every semisimple trivector is equivalent to
the linear combination  
\(
\lambda^{\rm ss}=\eta_1\lambda^{\rm ss}_1 + 
\eta_2 \lambda^{\rm ss}_2 + 
\eta_3 \lambda^{\rm ss}_3 + 
\eta_4 \lambda^{\rm ss}_4\;,
\)
of the trivectors
\bea
\lambda^{\rm ss}_1 &=& e_1 \wedge e_2 \wedge e_3 + e_4 \wedge e_5 \wedge e_6 
+ e_7 \wedge e_8 \wedge e_9\;,
\qquad 
\lambda^{\rm ss}_2 = e_1 \wedge e_4 \wedge e_7 + e_2 \wedge e_5 \wedge e_8 
+ e_3 \wedge e_6 \wedge e_9\;, 
\nonumber\\
\lambda^{\rm ss}_3 &=& e_1 \wedge e_5 \wedge e_9 + e_2 \wedge e_6 \wedge e_7 
+ e_3 \wedge e_4 \wedge e_8\;,
\qquad 
\lambda^{\rm ss}_4 = e_1 \wedge e_6 \wedge e_8 + e_2 \wedge e_4 \wedge e_9 
+ e_3 \wedge e_5 \wedge e_7\;,
\nonumber
\eea
where $\{ e_1 \cdots, e_9\}$ is a basis for $\C^9$ and
the coefficients $\eta_i$, $i=1,2,3,4$, are determined up to a linear
transformation by the Weyl group $W(\frak{e}_8)$ associated to the 
$\Z_3$-grading of $\frak{e}_8$
(see \cite{VE}). 
\footnote{Note that this
group is the huge Witting complex reflection group of order 155520.
Hence, the fact that the normal form is determined up the action of
the Witting group is not a trivial remark!}
We also know that this group is generated by complex reflections
with a parameter $\omega =e^{2\pi i/3}$, and the same result follows. 

\begin{prop}
The $SL(9)$-invariant configurations of junctions correspond to the admissible traces
(\ref{f3n}).
\label{adme8}
\end{prop}

  \subsection{Representations of 
the Lie algebra $\mathfrak{g}_0$ on $\otimes^3V$: The $E_6$ example}
  \label{e6 ex}

 We can embed $\mathfrak{e}_{6}$ in $\mathfrak{e}_{8}$ and
  compute the associated Lie commutators.  Recall that
  $\mathfrak{e}_{8} \supset \wedge^3 V$, where $V$ is a
  9-dimensional vector space, and that $\mathfrak{e}_{6} \supset V_{1}
  \otimes V_{2} \otimes V_{3}$, where $V_{i}$, $i = 1, 2, 3$, is a
  3-dimensional vector space, and similarly for the duals.

\vspace{3mm}
  Let $v_{i} \in V_{i}$ and $f_{j} \in V^{*}_{j}$, $i, j = 1, 2, 3$.
  Denote by $\overline{v}_{i} \in V$ ($\overline{f}_{j} \in V^{*}$)
  the extension of each vector (dual) to 9 dimensions by zero entries.
  (That is, $V_{1} \oplus V_{2} \oplus V_{3} \subset V$.) Then $V_{1}
  \otimes V_{2} \otimes V_{3} \subset \wedge^3 V$ by taking $v_{1}
  \otimes v_{2} \otimes v_{3} = \overline{v}_{1} \wedge
  \overline{v}_{2} \wedge \overline{v}_{3} \in \wedge^3 V$, and
  \begin{gather} [\overline{v}_{1} \wedge \overline{v}_{2} \wedge
    \overline{v}_{3}, \overline{f}_{1} \wedge \overline{f}_{2} \wedge
    \overline{f}_{3}] = -
    \begin{vmatrix}
      f_{1}(v_{1}) & 0 & 0 & \overline{f}_{1} \\
      0 & g_{2}(v_{2}) & 0 & \overline{g}_{2} \\
      0 & 0 & h_{3}(v_{3}) & \overline{h}_{3} \\
      \overline{v}_{1} & \overline{v}_{2} & \overline{v}_{3} &
      \frac{1}{3}I
    \end{vmatrix} ,
  \end{gather}
  from which we obtain
  \begin{gather} [v_{1} \otimes v_{2} \otimes v_{3}, f_{1} \otimes
    f_{2} \otimes f_{3}] = f_{1}(v_{1}) f_{2}(v_{2}) f_{3}(v_{3})
    \begin{bmatrix}
      \frac{1}{f_{1}(v_{1})} v_{1} \otimes f_{1} - \frac{1}{3}I \\
      \frac{1}{f_{2}(v_{2})} v_{2} \otimes f_{2} - \frac{1}{3}I \\
      \frac{1}{f_{3}(v_{3})} v_{3} \otimes f_{3} - \frac{1}{3}I
    \end{bmatrix}
    \in \mathfrak{sl}(V_{1}) \oplus \mathfrak{sl}(V_{2}) \oplus
    \mathfrak{sl}(V_{3}) .
    \label{e6}
  \end{gather}

\vspace{3mm}
   Given an order $r$  hypermatrix $A_{i_1\cdots i_r}$, $1 \leq i_j \leq n_j$,
  the hyperdeterminant of $A$ is invariant under $SL(n_1) \times \cdots \times
  SL(n_r)$ transformations.  In fact more generally it is relatively
  invariant under the action of $GL(n_1) \times \cdots \times
  GL(n_r)$ by \cite{GKZ} (Proposition 1.4, Chapter 14).
   This means that the hyperdeterminant of an order 3 hypermatrix $A_{i_1 i_2 i_3}$
  is invariant under $SL(n_1) \times SL(n_2) \times SL(n_3)$, or 
  $\frak{sl}(n_1) \oplus \frak{sl}(n_2) \oplus \frak{sl}(n_3)$ at the Lie algebra level.  
  Thus, 
  \begin{prop}
  For the $E_6$ model, $\frak{g}_0$ is the algebra leaving 
  invariant the hypermatrix factor. 
  \label{prop eee}
\end{prop}

Here a result similar to that of proposition \ref{adme8} also holds. 
However, to get the invariants explicitly requires calculations that
are outside the scope of this paper (we plan to get back to this
in the future). 
Semisimple and nilpotent elements, as well as the invariants 
are obtained in \cite{N}.
Note that the $\Z_3$-grading of $\mathfrak{e}_6$ and the computation of the
normal forms have been investigated in the context of quantum information
in reference \cite{BLTV}.

\subsection{Representations of 
the Lie algebra $\mathfrak{g}_0$ on $S^3V$: The $D_4$ example}
\label{d4 ex}
 Similarly, we have $\mathfrak{d}_{4} \subset \mathfrak{e}_{6}$
  by taking
\begin{gather}
  w_{1}w_{2}w_{3} = \sum_{\sigma \in S_{3}}
  \phi_{1}(w_{\sigma(i_{1})}) \wedge \phi_{2}(w_{\sigma(i_{2})})
  \wedge \phi_{3}(w_{\sigma(i_{3})}) \in S^3 W,
\end{gather} for isomorphisms $\phi_{i}: W \rightarrow V_{i}$, $i = 1,
2, 3$.  That is, $S^3 W = V_{1} \otimes V_{2} \otimes V_{3}$ for $W =
V_{1} = V_{2} = V_{3}$.  The same method applies to the dual.

\vspace{3mm}
Now we wish
  to express the Lie bracket \eqref{e6} for this algebra.  Denoting by
  $a_{j_{i}}'$ the vector $\phi_i(w_j)$ and $f_{j_{i}}'$ the vector
  $(\phi^{*})^{-1}_i(u^j)$ we have 
  \( [w_1w_2w_3, u^1u^2u^3] = \sum_{(j_{1},j_{2},j_{3})\in S_3}
  f_{j_{1}}'(a_{j_{1}}') f_{j_{2}}'(a_{j_{2}}') a_{j_{3}}' \otimes f_{j_{3}}' -
  \frac{1}{3}I\;. 
  \)

\vspace{3mm}
 We may express the action $[\mathfrak{g}_0,\mathfrak{g}_1]
  \rightarrow \mathfrak{g}_1$ in terms of the matrix $M\in
  \mathfrak{sl}(W)$ and $\l\in S^3W$.  Given $\l=u\otimes v\otimes w,$
  we have the transformed $\l'=Mu\otimes Mv\otimes Mw,$ or using the
  notation from before \(
  \l'=(M,M,M)\cdot \l\;.
  \label{MMM}
  \) 
  The action is
  similarly defined for $\mathfrak{g}_{-1}$.

\vspace{3mm}
 A sufficient condition for $\l$ to be left invariant is that
  $u$, $v$, and $w$ are eigenvectors of $M$ with eigenvalue 1, or
  \(\det(M-I)=0.\)
From Proposition \ref{discrim} we have that the
transformation formula for the  hyperdeterminant \(\Delta(\l')=\det(M)^9 \Delta(\l),\) so for
  invariance of $\l$ we must have $\det(M)^9=1$. Therefore, we get
  
  \begin{prop}
  A state in a junction in the $\frak{d}_4$ model 
  is invariant if $\det (M)^9=1$. 
  \end{prop}
 
  This can happen, for example, for $M=I e^{2\pi i/9}$, i.e. a 9th root of unity. 
  
  \medskip
\noindent  { \bf Remark.}
   Again, a result similar to that of proposition \ref{adme8} also holds here. 
   However, as we noted right after proposition \ref{prop eee}, we leave
   the explicit computation of the invariants for a future treatment.
  
  \subsection{The non-simply laced Lie algebras: Types $F_4$ and $G_2$}
  \label{f4 ex}
  The non-simply laced exceptional groups do not include a third
  (antisymmetric, symmetric or tensor) power in their graded decomposition.
  However, there is a 3-tensor symmetric on two indices in the case of 
  $F_4$, and an extra 3-form is involved in the case of $G_2$.
  
  \vspace{3mm}
\noindent{\bf  Representations of 
the Lie algebra $\mathfrak{g}_0$ on $S^2V\otimes V$: The $F_4$ example.}
   The Lie algebra $\frak{f}_4$ of the Lie group $F_4$ admits
   the $\Z_3$-graded decomposition  
  \( \mathfrak{f}_4 = (S^2V_1 ^{*} \otimes V_2^{*}) \oplus 
  (\mathfrak{sl}(V_1) \oplus \mathfrak{sl}(V_2)) \oplus
  (S^2V_1 \otimes V_2)\;, ~~~~ \dim V_i=3.
  \)
  We see that the factor $S^2 V_1 \otimes V_2$ is the part of 
  $V_1 \otimes V_3 \otimes V_2$, where two vector spaces
  $V_1$ and $V_2$ are identified. The Lie algebra 
  $\frak{sl}(V_1) \oplus \frak{sl}(V_2)$ can be embedded in 
  $\frak{sl}(V)$ so that any element $(X, Y)$ in the former 
  corresponds to the block-diagonal matrix with blocks 
  $X, X, Y$. This allows for computation of the commutators in the
  algebra. Semisimple and nilpotent elements, as well as 
  invariants can be found (also for $E_6$ and $D_4$) in 
  \cite{AN}.
  
 
  \vspace{3mm}
  \noindent {\bf Remark.} Notice that what appears here as a summand is 
  a symmetric analog of the degree three element that is antisymmetric 
  on the first two indices which appears in the $GL(n)$-decomposition 
  of $V^{\otimes 3}$, for instance in expression (\ref{s21}).

\vspace{3mm}  
\noindent {\bf Invariant 3-forms and the $G_2$ case.}
   $G_2$ does not admit a cubic factor in its graded 
  Lie algebra decomposition. The dimension of the Lie algebra
  ${\sf g}_2={\rm Lie}(G_2)$ is 
  too small to admit such a factor, but it admits the decomposition
  \(
 {\sf{g}}_2(\C)= {\rm Lie}(G_2)=V^* \oplus \frak{sl}(V) \oplus V\;,~~~~~\dim_{\C} V=3.
  \label{g2 dec}
  \)
  By duality, the factors $\frak{g}_{-1}$ and $\frak{g}_1$ can alternatively 
  be taken to 
  be $\wedge^2V^*$ and $\wedge^2V$, respectively.
  The real version admits a similar decomposition but with a real vector
  space dimension four. This 
   is used in \cite{G2} to give a superalgebra structure on ${\sf g}_2(\R)$
  and to make connection to symmetries of multiple membranes and 
  Lie 3-algebras. 
  
  \vspace{3mm}
  We see from (\ref{g2 dec}) that in this case a 3-form, 
  for instance, would have to be an additional piece of data, 
  i.e. $\lambda \notin \frak{g}_1$.
  Consider invertible complex linear transformations $S$ on a 
  3-form $\lambda$ on a complex 7-dimensional vector space
  $V$ such that 
  \(
  \lambda(M \cdot, M\cdot, M\cdot)=\lambda (\cdot, \cdot, \cdot)\;.
  \) 
The group of such $M$ is $G_2 \times \Z_3$ \cite{Herz}.

\vspace{3mm}
Given the above, the transformation (\ref{MMM}), and Proposition
\ref{discrim},  
we therefore have
\begin{prop}
The states of a 3-junction, 
represented by a three form on a complex  seven dimensional vector space $V$, 
 are invariant under 
the algebra ${\sf{g}}_2$ or the group $G_2 \times \Z_3$. 
\end{prop}

In the non-simply laces case, a result similar to that of proposition \ref{adme8} also holds.

\section{Further Applications and Extensions}

\subsection{Symmetry of dimensionally-reduced supergravity}
\label{sugra}

In \cite{CJLP} it was shown that the underlying algebras 
for  all the $D$-dimensional maximal supergravities
that come from eleven dimensions
 are deformations of $\mathcal{G}\oplus_s \mathcal{G}^*$, where $\mathcal{G}$ 
 itself is the
semi-direct sum of the Borel subalgebra of the superalgebra
$\frak{sl}(11-D|1)$ and a rank-3 tensor representation, and $\mathcal{G}^*$ is the
coadjoint representation of $\mathcal{G}$.
 The fields
coming from the 3-form potential in $D=11$ form a linear graded
antisymmetric 3-tensor representation of $\frak{sl}(n|1)$.  The algebra
$\mathcal{G}$ for $D$-dimensional supergravity can be denoted by
\(
\mathcal{G}={\frak sl}_+(n|1) \oplus_s (\wedge V)^3\ .
\label{Ggroup}
\)
with $V$ the appropriate fundamental representation, and 
$\frak{sl}_+(n|1)$ is 
the Borel subalgebra
of the superalgebra $\frak{sl}(n|1)$. 

\vspace{3mm}
In the special case of a reduction to
$D=3$ dimensions, the obvious $\frak{gl}(n,\R)$ symmetry 
from the dimensional reduction on an $n$-torus 
can be enlarged to the bosonic algebra ${\frak sl}(n+1,\R)$
rather than the superalgebra ${\frak sl}_+(n|1)$.  
In the case of the doubled
system of equations for maximal supergravity in $D=11-n$ dimensions,
the global part of the gauge field preserving symmetry is $\frak{e}_n^+$,
the Borel subalgebra of the algebra $\frak{e}_n$.
In \cite{CJLP}
it was expected that the doubled formalism should
be invariant under the full global $\frak{e}_n$ algebra.
Here we provide a proof of that for the case $n=8$. 

\vspace{3mm}
This is actually straightforward in our setting. For $n=8$, 
the enhanced algebra from the 8-torus will be 
$\frak{sl}(V)$, with ${\rm dim}(V)=9$ (rather than 
${\rm dim}(V)=8$). 
The algebras $\mathcal{G}$ and $\mathcal{G}^*$ are then
 \bea
 \mathcal{G}&=& \frak{sl}(V) \oplus \wedge^3 V\;,
 \nonumber\\
 \mathcal{G}^*&=& \frak{sl}(V^*) \oplus \wedge^3 V^*\;.
\eea
Now forming the semidirect sum gives
\(
\mathcal{G}\oplus_s \mathcal{G}^*=
\wedge V^* \oplus \frak{sl}(V) \oplus \wedge^3 V
\)
But this is exactly the $\Z_3$-graded model of $\frak{e}_8$ 
(see Proposition \ref{prop case}). Thus we immediately have
the following

 \begin{theorem}
 In the doubled formalism, the symmetry of gauge fields resulting from the 
 dimensional reduction of eleven-dimensional 
 supergravity on the 8-torus is $\frak{e}_8$.
 \end{theorem}

Note that at the level of Lie algebras, we take $\frak{e}_8$ to be 
a real form of the corresponding complex Lie algebra.

\subsection{Valued-ness of the fields}
Our discussion suggests that the fields in the adjoint 
representation of $\frak{sl}$, $\frak{so}$ and $\frak{sp}$,
respectively, would be replaced by fields in the corresponding 
degree three antisymmetric, tensor, and symmetric powers
\bea
A^{[ij]}~e_i \wedge e_j \qquad &\leadsto& \qquad  
A^{[ijk]}~e_i \wedge e_j\wedge e_k\;,
\nonumber\\
A^{ij}~e_i \otimes e_j \qquad &\leadsto& \qquad
A^{ijk}~e_i \otimes e_j \otimes e_k\;,
\nonumber\\
A^{(ij)}~e_i \odot  e_j \qquad &\leadsto& \qquad
A^{(ijk)}~e_i \odot  e_j\odot e_k\;.
\eea
What possible combinations of the wedge $\wedge$, tensor
$\otimes$, and
symmetric $\odot$ products can occur, i.e. which indices 
$i, j, k$ are admissible? This of course depends on the
Lie algebra $\frak{g}$. In general, there is a Jordan 
decomposition of such degree three tensors into 
a semisimple part and a nilpotent part and the admissible
tensors are known (see the discussion in section \ref{tr}
and section \ref{e8 ex}).

\subsection{Higher $m$-vectors}
\label{duality symmetry}
Here we provide alternatives to the models presented in 
sections \ref{e8 ex} and \ref{e6 ex}. We have seen 
from equation (\ref{cc}) that $\wedge^3 V$ does not 
close on itself but rather on the dual $\wedge^3V^*$ 
(cf. the remarks at the end of section \ref{e8 ex}).
Here we describe a model in which the closure is
on $\wedge^6 V$, i.e. for $\lambda_1, \lambda_2 \in 
\wedge^3 V$ we have the commutator
\(
\left[\lambda_1~,~\lambda_2 \right]=\l_1 \wedge \l_2\;.
\label{ll}
\)
Thus we seek a graded Lie algebra decomposition which includes 
$\wedge^6V$ as a summand. 
Then we would have the following extra cases (see \cite{OV}):
\begin{enumerate}
\item $\frak{e}_6=\wedge^6 V^* \oplus \wedge^3 V^* \oplus
\frak{gl}(V) \oplus \wedge^3 V \oplus \wedge^6 V$, ~~~~~~~~~$\dim (V)=6$.

\item $\frak{e}_7=\wedge^6 V^* \oplus \wedge^3V^* \oplus \frak{gl}(V) \oplus \wedge^3 V \oplus \wedge^6 V$,
~~~~~~~~~$\dim (V)=7$.

\item $\frak{e}_8= (V^* \otimes \wedge^8 V^*)
\oplus \wedge^6 V^* \oplus \wedge^3 V^* \oplus \frak{gl}(V)
\oplus \wedge^3 V \oplus \wedge^6 V \oplus 
(V\otimes \wedge^8 V)$, $\dim (V)=8$.

\end{enumerate}
In addition to (\ref{ll}) there are other brackets corresponding to 
each pair of summands in the above decompositions of 
$\frak{e}_i$, $i=6,7,8$. Most relevant for us is the bracket 
of $\lambda \in \wedge^3 V*$ and $C \in \wedge^6 V$
\(
\left[ \lambda^*~,~C\right] =\frac{1}{6}\lambda^{*l_1 l_2 l_3}
C_{l_1 l_2 l_3ijk}\;.
\label{lc}
\)
The bracket between $\lambda \in \wedge^3 V$ and 
$C^* \in \wedge^6 V^*$ is obtained from 
(\ref{lc}) by simply raising all lower indices and lowering all
upper indices. The complete brackets are given for instance 
 in \cite{OV}.

\vspace{3mm}
\noindent {\bf Remarks.} 
{\bf 1.} Because of the identity $[\frak{g}_0~,~\frak{g}_i]\subset \frak{g}_i$
we get representations of the algebra $\frak{g}_0=\frak{gl}(V)$ on $\wedge^6V$.
As in section \ref{rep groups} we also get representations of the 
corresponding general linear groups on $\wedge^6V$.

\noindent {\bf 2.}
 One can in principle consider the action of $GL(V)$ 
 which breaks $V^{\otimes 6}$ into a direct sum of 
$GL(V)$-modules which include
\(
V^{\otimes 6} \supset \wedge^6V \oplus S^6V\;.
\)
This is a special cases of the more general action of 
$GL(V) \times \Sigma_n$ on $V^{\otimes n}$ leading to the
canonical isomorphism 
$V^{\otimes n} \cong \bigoplus S_{\rho} (V) \otimes V_{\rho}$,
where the sum is over all partitions $\rho$ of $n$ into 
at most ${\rm dim} V$ parts, and $S_{\rho}(V)$ is the (image of
the) Schur functor, i.e. the image 
$S_{\rho}(V)={\rm Im} (c_{\rho}: V^{\otimes n} \to V^{\otimes n})$
of the Young symmetrizer
 $c_{\rho}\in \C\Sigma_n$ 
(see \cite{FH}).

Thus we could have posed the question as that of  
 seeking graded Lie algebra decompositions that include (one of)
the summands $\otimes^6V$ or $\wedge^6V$ or $S^6V$.
The question in the antisymmetric cases is provided by the 
above three cases of exceptional Lie algebras of type $E$. 

\noindent {\bf 3.} What does (\ref{ll}) mean in terms of 
states and configurations? It represents a composite of two
3-junctions that are not joined or do not intersect. 

\noindent {\bf 4.} The bracket (\ref{lc}) represents the contraction between 
a dual 3-junction state and a composite of two 3-junction states, giving
rise to a single 3-junction state. This is a degree three analog of the
contraction of a degree two tensor by a metric. 

\noindent {\bf 5.} The degree six factor suggests the field coupling to 
the fivebrane. This forms part of the discussion in the next section.



\subsection{Generalized Born-Infeld for membranes and fivebranes?}
\label{GBI}

\noindent {\bf D-branes.}
The dynamics of D-$p$-branes, with $d=p+1$ spacetime dimensions,
is described in part by the Born-Infeld action of nonlinear electrodynamics.
This can be seen from the sigma model approach \cite{Le}  
or using path integrals \cite{Ts}. The action is given by
\(
S_d= \int d^d x e^{-\phi} \sqrt{\det \left(g_{mn} + \mathcal{F}_{mn} \right)}\;,
\label{dbi}
\)
where $\mathcal{F}_{mn}=F_{mn} - B_{mn}$ is the difference (or, alternatively, 
sum) of the 
components of the curvature 2-form $F_2$ 
of the $U(1)$ bundle and the $B$-field $B_2$. 

\vspace{3mm}
\noindent {\bf The membrane.}
 The fields on an open membrane include a 3-form field strength $F_3$, whose 
 potential is a 2-form $A_2$ on the boundary. The 3-form 
can be combined with the pullback of the background $C$-field $C_3$ to form the 
shifted field
\(
H_3=F_3 - C_3\;.
\label{HF}
\)
This is a higher degree analog of the gauge invariant combination 
$F_2 -B_2$ for the open string, where $F_2$ is the curvature of the $U(1)$ 
bundle and $B_2$ is the connection on a gerbe. 

\vspace{3mm}
\noindent {\bf The fivebrane.} The topological part of the 
worldvolume action involve combinations of 
the expression (\ref{HF}) as well as \cite{Wi} \cite{T}
\(
H_6=F_6 + C_6 + {\rm composite}\;.
\label{H6}
\)
 There exist proposed extensions that involves metric-dependent terms. 
 One is the PST action which has an auxiliary scalar $a$ and 
a dual field $H_2=*_6 (da\wedge H_3)$ in six dimensions. The gauge-invariant 
action involves \cite{PST}
\(
S_{PST}\supset \int
 \left( -d^6\sigma \sqrt{\det (g + H_2)} + 
 (C_6 + \frac{1}{2}F_3 \wedge C_3)
 \right)
 \;.
 \label{PST}
\)
The dimensional reduction reproduces the action of the D4-brane via the 
identifications $F_3 \to (F_3, F_2)$, $C_3 \to (C_3, B_2)$, and $C_6 \to C_5$.

\vspace{3mm}
\noindent {\bf Higher `gerby' Born-Infeld.}
The boundaries of the membrane -- which can end on M5-branes--
 are strings and hence carry 
not gauge but gerbe degrees of freedom. 
Gerbes model higher form electrodynamics so that it is natural to 
ask for a nonlinear  version of such a higher form. Thus, we propose a higher
generalization of Born-Infeld action to accommodate degree three and degree
six field strengths corresponding to the membrane and the fivebrane, respectively.
As recalled in section \ref{hh}, what replaces the determinant det is naturally the {\it hyperdeterminant}
${\mathcal{D}}{\rm et}$. 
Furthermore, there is 
no obvious metric part in this case (unless we consider the idea of the dual 
of the graviton; see \cite{Cu} \cite{West}). Thus, a generalization of  
the action (\ref{dbi}) and (\ref{PST}), without the gravity part, 
would involve a scalar built out of the fields 
(\ref{HF}) and (\ref{H6}) for the case of the membrane and fivebrane,
respectively. 

\vspace{3mm}
We consider the antisymmetric tensor fields 
$H_3$ and $H_6$ as hypermatrices of the form 
\begin{itemize}
\item {\it For membrane:} 
$H_3=(H_{ijk})_{1 \leq i,j,k \leq 3}$,

\item {\it For fivebrane:}
\begin{itemize}
\item
$H_3=(H_{ijk})_{1 \leq i,j,k \leq 6}$,
\item $H_6=(H_{i_1 i_2 \cdots i_6})_{1 \leq i_k \leq 6}$,  \hspace{5mm}$k=1, \cdots, 6$.
\end{itemize}
\end{itemize}

The desired action will involve a square root of a generalization of the
determinant. In the case of an antisymmetric matrix, this has an 
interesting description in terms of a Pfaffian, which is the 
 `square root' of the determinant of an antisymmetric matrix. 
 In fact, a Pfaffian can
be described in several ways, all of which turn out to be equivalent. 

\vspace{3mm}
The analog of an antisymmetric matrix can be defined as follows. 
A $k$-dimensional alternating tensor $A$ of order $n$ can be defined as 
a function $A$ on the product set $\{1, \cdots, n\}^n$ such that
\(
A(i_1, \cdots, i_k)={\rm sign} (\sigma) A(i_{\sigma(1)}, \cdots, 
i_{\sigma (k)})
\)
for any permutation $\sigma \in \Sigma_k$ and 
$1 \leq i_1, \cdots, i_k \leq n$. 

\vspace{3mm}
The higher degree analog of the Pfaffian will be the {\it hyperpfaffian},
which plays the analogous role for the hyperdeterminant of an 
alternating tensor as the Pfaffian plays for the determinant of an
antisymmetric tensor. Like the Pfaffian, there are several ways of 
defining the hyperpfaffian. However, in contrast to the pfaffian, 
those definitions are not all equivalent (for a discussion on this 
see the first section in \cite{Red}). Some definitions of the Pfaffian
are, like the hyperdeterminant (see section \ref{hh}), 
the zero polynomial for the case when $k$ is odd. This will not 
be useful for us because we are seeking an expression involving 
$H_3$, i.e. for $k=3$. Luckily, there is a definition that works for both
even as well as odd $k$ \cite{LT} and which is the one we will 
follow. 

\vspace{3mm}
Let $\Sigma_{km, k} \subseteq \Sigma_{km}$ be the set of permutations 
$\sigma$ such that 
$$
\sigma (ki +j) < \sigma (ki + j +1)~~{\rm and}~~ \sigma (ki+1)< \sigma (k(i+1)=1)
~~{\rm for~all}~0\leq i < m~~{\rm and}~1 \leq j <k.
$$
Then for a $k$-dimensional alternating tensor $A$ of order $km$, the
{\it hyperpfaffian} of $A$ is defined to be \cite{LT}
\footnote{There is an earlier definition of the hyperpfaffian in \cite{Ba}, but 
that definition matches only for the even case.}
\(
{\it Pf}_k(A)=\sum_{\sigma \in \Sigma_{mk,k}} {\rm sign}(\sigma)
\prod_{i=0}^{m-1}
A(\sigma (ki+1), \cdots, \sigma (ki+k))\;.
\label{hpf}
\)
For $k=2$, this reproduces the formula for the Pfaffian as follows 
(see \cite{Red}). Define $\mathcal{S}_{2n} \subseteq \Sigma_{2n}$ to be the
set of all $\sigma \in \Sigma_{2n}$ such that $\sigma (i) < \sigma (i+1)$
and $\sigma (i) < \sigma (i+2)$ for all odd $i$. Then for a $2n \times 2n$
antisymmetric matrix $A$ the Pfaffian is 
\(
{\rm Pf} (A)= \sum_{\sigma \in \mathcal{S}_{2n}}
{\rm sign}(\sigma) 
\prod_{i=0}^{n-1} 
A(\sigma (2i+1), \sigma (2i+2))\;,
\)
which indeed coincides with (\ref{hpf}) for $k=2$.


\vspace{3mm}
We notice from (\ref{hpf}) that the order of the antisymmetric tensor
should be a nontrivial multiple of its dimension. This means that 
the case for $H_3$ on the membrane worldvolume cannot be described
by expression (\ref{hpf}), whereas both $H_3$ and $H_6$ on the fivebrane
worldvolume do.
The proposed action for the fivebrane would then contain
\(
S_{M5} \supset \int {\rm Pf} (H_6)\;,
\label{DBI6}
\)
where $H_6$ has expression (\ref{H6}). 
For the membrane, while we cannot write a similar expression using the same
definition for the hyperpfaffian, we expect something analogous to occur once
a convenient definition for the hypepfaffian is obtained which can be adapted 
for the case when the dimension of the tensor is equal to its rank. 

\begin{Proposal}
{\it M5-branes (and M2-branes) can be described (in part) by 
the generalized Born-Infeld action (\ref{DBI6}) (and a similar action 
for the M2-brane)}.
\end{Proposal}

\vspace{3mm}

\begin{figure}
\centering
\includegraphics[scale=.8]{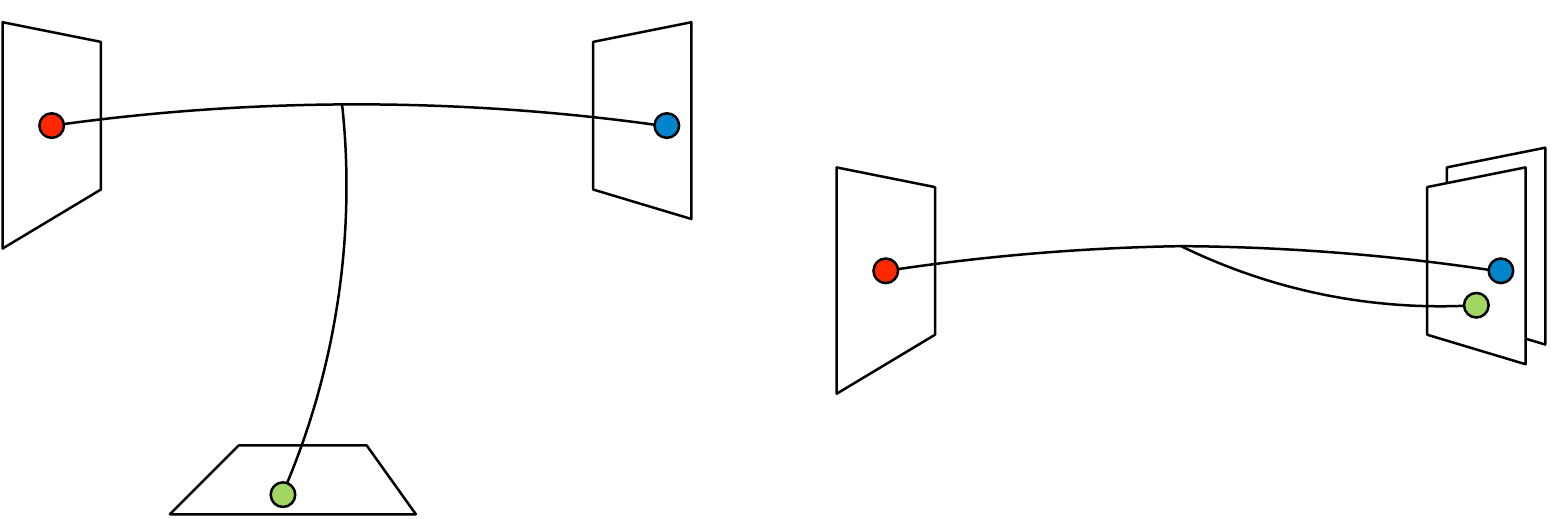}
\caption{Junctions reproduce two-sided open strings when two of the
  three segments overlap.}
  \label{correspondence}
\end{figure}

\noindent  {\bf Correspondence with the string/D-brane case.} 
The determinant is 
part of the formula for the hyperdeterminant. 
We consider $F$ to be sitting inside $H$ as a slab, so that we get 
a matrix of we start with a hypermatrix all of whose slabs in one
direction are the same, i.e. if the hypermatrix is a stack of 
identical slabs. By slab operations, this is equivalent to 
a hypermatrix with all zero entries except in one full slab. 
For a visualization see figure  \ref{correspondence}.

\subsection{Final Remarks}

\vspace{3mm}
\noindent{\bf 1. Relation to Kac-Moody Algebras.}
We discussed a duality-symmetric model of the $E$-series in section
\ref{duality symmetry}. In fact $E_8$ is duality-symmetric in a
different setting.  This is one main aspect of the $E_9$, $E_{10}$ and
$E_{11}$ models aiming to describe ten- and eleven-dimensional
supergravity and M-theory-- see \cite{West}.

\vspace{3mm}
There is a correspondence \cite{Kac} between a $\Z_m$-graded 
Lie algebra $\frak{g}$ with its `covering' infinite-dimensional $\Z$-graded
Lie algebra
\(
\widehat{\frak{g}}=\sum_{k \in \Z} \widehat{\frak{g}}_k \subset \C [t,t^{-1}]\otimes \frak{g}\;,
~~~~~~ \widehat{\frak{g}}_k=t^k \frak{g}_k\;,
\)
where $\frak{g}_k$ denotes the grading subspace of $\frak{g}$ whose index is the
residue class $k$ modulo $m$. 
So obviously any $\frak{g}_k \subset \frak{g}$ will also be a summand 
in $\widehat{\frak{g}}$.

\vspace{3mm}
The algebra $\frak{g}$ is obtained, as an algebra over $\C$,
 from $\widehat{g}$ by factoring the ideal $(u-1)\widehat{g}$, where the 
 multiplication $u$ is defined by the formula $ux=t^m x$, $x \in \widehat{g}$,
 which make $\widehat{g}$ a finite-dimensional 
 $\C[u,u^{-1}]$ algebra.


\vspace{3mm}
The models we have seen in this paper use finite-dimensional 
-- and in fact relatively low-dimensional -- vector spaces. On 
the other hand we would be interested in the large $N$ limit,
which thus cannot be immediately seen in such models. It might 
be possible that embedding in a Kac-Moody algebra might allow
for this possibility, but we will not discuss this further in the current
paper.

\vspace{3mm}
\noindent{\bf 2. Relation to 3-algebras.}
In the main part of this paper we focused on keeping Lie algebras
in the discussion. 
The Lie bracket on a Lie algebra $\frak{g}$ is defined as a map 
$[~,~]: \wedge^2 \frak{g} \to \frak{g}$. There has been very 
interesting recent activity (starting with \cite{BL} and \cite{G}) 
on modeling multiple M-branes using
Lie 3-algebras with bracket $[~,~,~]: \wedge^3 \frak{g} \to \frak{g}$.

\vspace{3mm}
In \cite{DFMR} a Lie-algebraic origin of certain metric
3-algebras is provided. In
particular, it is proved that certain metric 3-algebras 
correspond to pairs $(g, V )$ consisting of a metric Lie algebra $g<\frak{so}(V)$,
$[g,g]\subset g$, and a real faithful orthogonal
representation $V$. In this paper we kept working with Lie algebras 
(justified by \cite{DFMR})
and used third powers of $V$ instead of $V$ itself. 
Thus, we have taken a 
different path from the ones in the above cited works.
Hence the work in this paper will not directly connect to Lie 3-algebras but 
could be seen as complimentary. Further work might require
higher algebras as in \cite{SSS1}.



%

%

%

\vspace{3mm}
We have presented models that capture some aspects of 
 the description of 3-junctions
which introduces hypermatrices and their hyperdeterminants. 
This made natural and novel connection to exceptional Lie algebras.  
However, as we discussed throughout the paper, there are many unanswered
questions. Our treatment has been mostly formal, and  further physical 
arguments would be needed to tell how the physics of D-branes 
would favor a model over the other. Furthermore, a more refined 
mathematical discussion might be needed. 
We hope to address such matters in a future project to 
at a more final answer. The full answer is likely to go beyond usual
(non)linear algebra.

     \bigskip\bigskip
\noindent
{\bf \large Acknowledgements}

\vspace{2mm}
\noindent The authors would like to thank the Mathematics Department at Yale
University, especially Mikhail Kapranov and Yair Minsky, for support and encouragement 
in the Research Experience for Undergraduates (REU) on {\it Nonlinear Algebra and Hyperdeterminants} in summer of 2009. This paper builds directly on results from that 
program, ran by H. S., 
and in which Y. F. (a  sophomore), S. L. (a freshman) and D. T. (a freshman) were 
participating students. H. S. thanks Mikhail Kapranov for useful comments. 
The authors are grateful to the (anonymous) referees for very useful 
suggestions.

\end{document}